\pdfoutput=1
\documentclass[10pt]{amsart}
\usepackage[T1]{fontenc}
\usepackage{lmodern}
\usepackage{amsmath,amssymb,mathrsfs,mathtools,enumitem,booktabs,graphicx,microtype,xcolor}
\usepackage[colorlinks=true,linkcolor=blue!50!black,citecolor=blue!50!black,urlcolor=blue!50!black]{hyperref}
\allowdisplaybreaks[2]
\numberwithin{equation}{section}

\theoremstyle{plain}
\newtheorem{theorem}{Theorem}[section]
\newtheorem{proposition}[theorem]{Proposition}
\newtheorem{lemma}[theorem]{Lemma}
\newtheorem{corollary}[theorem]{Corollary}
\theoremstyle{remark}
\newtheorem{remark}[theorem]{Remark}

\newcommand{\R}{\mathbb R}
\newcommand{\E}{\mathbb E}
\newcommand{\PP}{\mathbb P}
\newcommand{\C}{\mathcal C}
\newcommand{\D}{\mathcal D}
\newcommand{\G}{\Gamma}
\newcommand{\op}{\mathscr P}
\newcommand{\Lh}{\mathcal L}
\newcommand{\loc}{\mathrm{loc}}
\newcommand{\tr}{\operatorname{tr}}
\newcommand{\supp}{\operatorname{supp}}
\newcommand{\norm}[1]{\left\lVert#1\right\rVert}
\newcommand{\abs}[1]{\left\lvert#1\right\rvert}

\title[Zero-variance geometry for Heston American puts]{Interior smoothness and zero-variance geometry\\ for finite-maturity Heston American puts}

\author{Junkee Jeon}
\address{Department of Applied Mathematics, Kyung Hee University, Yongin 17104, Republic of Korea}
\email{junkeejeon@khu.ac.kr}

\author{Takwon Kim}
\address{School of Mathematics, Statistics and Data Science, Sungshin Women's University, Seoul 02844, Republic of Korea}
\email{takwon@sungshin.ac.kr}

\author{Jehan Oh}
\address{Department of Mathematics, Kyungpook National University, Daegu 41566, Republic of Korea}
\email{jehan.oh@knu.ac.kr}


\begin{document}

\begin{abstract}
We study finite-maturity American puts in the Heston model without the
Feller condition. At positive variance, the exercise boundary is smooth
and strictly decreasing in remaining maturity and variance. At zero
variance, the endpoint is locally Lipschitz in remaining maturity, the
boundary moves with bounded speed in maturity uniformly down to zero
variance, and it extends continuously to the axis. Each
boundary slice at positive maturity has a regular $C^1$ closure tangent
to the zero-variance axis, its stock threshold has variance
derivative tending to minus infinity, and both partial derivatives of
the inverse boundary graph vanish along the slice. The continuation
premium at the axis contact stock is of order variance, and as a
consequence smooth fit fails in the variance direction at the axis. At
expiry the whole boundary, including its axis edge, converges to the
strike. We give quantitative boundary gap bounds, an exact limiting
weighted boundary Hessian, and vertical identities retaining the
maturity term. Exact boundary gap equivalents remain open.
\end{abstract}

\maketitle

\section{Introduction and main results}

The exercise boundary of an American put under stochastic volatility is a
surface in time, stock price and variance. Knowing that it exists and that it
is monotone does not tell us whether it is continuous, and knowing that the
value function satisfies smooth fit does not tell us whether the surface is
differentiable. The Heston generator adds one more difficulty, because it is
uniformly parabolic on compact subsets of positive variance but degenerates
at zero variance.

In this paper we prove that the whole interior exercise boundary of the
finite-maturity Heston put is smooth. Three ingredients carry the argument.
Monotonicity of the price in the stock and in the variance shows that every
contact point carries a fixed spatial orthant of the contact set. An
elementary lifting embeds the discounted equation, drift included, into
exactly the class of time-monotone parabolic obstacle problems treated by
Audrito and Kukuljan \cite{AK}. The contact orthant then excludes every
singular quadratic blow-up. A separate Hopf argument shows that the resulting
smooth hypersurface is a smooth graph over the stock direction, and this last
step cannot be replaced by monotonicity alone.

For the one-dimensional Black and Scholes put, the variational inequality
formulation goes back to Jaillet, Lamberton and Lapeyre \cite{JLL}, and the
early exercise premium and the integral equation for the boundary to Kim
\cite{Kim}, Jacka \cite{Jacka} and Myneni \cite{Myneni}. Chen and Chadam
\cite{CC} analyzed the smoothness of the put boundary and its behavior near
expiry, Chen, Chadam, Jiang and Zheng \cite{CCJZ} proved its convexity,
Peskir \cite{Peskir05} settled uniqueness for the integral equation, and the
monographs of Peskir and Shiryaev \cite{PS} and Detemple \cite{Detemple} and
the survey of Ekstr\"om \cite{Ekstrom} present smooth fit, the free-boundary
equation, and the monotonicity and convexity properties of American prices.
In genuinely two-dimensional finite-horizon problems, continuity of stopping
boundaries and smooth fit have been obtained with probabilistic tools by
Peskir \cite{Peskir19}, by De Angelis and Peskir \cite{DP20}, and by Cai, De
Angelis and Palczewski \cite{CDP} for the put with a stochastic interest rate.
Higher regularity of a boundary that depends on two state variables and on
time requires the local theory of parabolic obstacle problems, and this is
the route taken here.

For stochastic volatility models, Touzi \cite{Touzi} examined the dependence
of the exercise boundary on the volatility state under a uniform ellipticity
assumption, which the Heston model does not satisfy, Assing, Jacka and Ocejo
\cite{AJO} proved monotonicity of the price in the volatility for a class
containing the Heston model under the Feller condition, and Chen, Jin and
Song \cite{CJS} studied continuity and monotonicity of the boundary for a
class of stochastic volatility models by PDE methods. Lamberton and Terenzi
\cite{LTvi} gave a variational formulation of the American price in weighted
Sobolev spaces and in \cite{LTprop} established, without the Feller
condition, continuity of the price, monotonicity in the variance and in
time, convexity in the stock, a positive exercise threshold, and the weak
smooth fit $P_s=-1$ along the boundary, whereas the variance derivative was
shown to vanish on the boundary only under the Feller condition. Nguyen and
Chau \cite{NC} studied interior Sobolev regularity and smooth fit for the
finite-maturity problem under that condition. Here the local Sobolev
argument is given independently, and the smoothness and strict monotonicity
of the interior exercise surface and the full smooth-fit identities are
obtained without the Feller condition.

On the side of obstacle problems, the regularity of the free boundary near
regular points goes back to Caffarelli \cite{Caf77,Caf98} in the elliptic
case and to Friedman \cite{Friedman75} for one-dimensional parabolic
variational inequalities, the parabolic theory of Caffarelli, Petrosyan and
Shahgholian \cite{CPS} classifies blow-ups at free-boundary points, and the
monograph of Petrosyan, Shahgholian and Uraltseva \cite{PSU} and the survey
of Petrosyan and Shahgholian \cite{PS07} collect the theory and its
applications to finance. Blanchet, Dolbeault and Monneau \cite{BDM} proved
continuity of the time derivative for parabolic obstacle problems with
variable coefficients, and Laurence and Salsa \cite{LS} obtained smooth
exercise boundaries for options on several geometric Brownian assets, where
the transformed operator has constant coefficients. The regular-part theorem
of Audrito and Kukuljan \cite{AK}, which rests on the higher-order boundary
Harnack inequality of Kukuljan \cite{Kuk}, is the mechanism we use after the
lifting and after the exclusion of singular points. In the degenerate
direction, Daskalopoulos and Feehan \cite{DF} proved weighted optimal
regularity for a degenerate elliptic obstacle problem, Feehan and Pop
\cite{FP} developed Schauder theory for linear degenerate parabolic
equations, and Wang and Zhang \cite{WZ} obtained $C^{1,\alpha}$ estimates
for degenerate parabolic equations of Heston type. The zero-variance
estimates below are obtained directly from the stochastic model and from
interior rescaling arguments, without these weighted estimates at the
degenerate contact point.

The analysis at zero variance requires a separate argument. The price
increment caused by an extension of maturity is controlled through the
expected stock local time at the strike, uniformly as the initial variance
vanishes, and dynamic programming then gives a linear upper bound for the
premium at the axis contact point. An intrinsic parabolic rescaling gives the
matching lower bound, uniformly on compact maturity intervals, and an entry
argument at small times excludes a boundary gap of order variance. A strict
waiting loss estimate inside the stopping interval controls the displacement
of the threshold under a maturity extension at every variance level
including zero. It excludes jumps, gives a Lipschitz modulus for the axis
endpoint, and bounds the maturity derivative of the boundary uniformly down
to the axis. Moving contact rescalings then prove tangency and a regular
$C^1$ closure of each spatial boundary slice, along which both partial
derivatives of the inverse boundary graph vanish. The linear lower bound for
the premium shows that the variance derivative of the price cannot be
continuous at the axis contact point, so smooth fit fails in the degenerate
direction there, and a weak formulation argument shows that the boundary
converges to the strike at expiry, uniformly on bounded variance sets and
including the axis. Exact boundary gap equivalents, the differentiability of
the axis endpoint, and the continuity of the stock derivative of the price at
the axis require additional estimates, and we state precisely what is
missing.

Throughout the paper we fix parameters
\begin{equation}\label{assumptions}
 r>0,\qquad 0\le q\le r,\qquad
 \kappa>0,\quad \theta>0,\quad \sigma>0,\quad K>0,\quad -1<\rho<1.
\end{equation}
In the Heston model \cite{Heston}, under the pricing measure the stock and the variance satisfy
\begin{equation}\label{sde}
 \begin{aligned}
 dS_u&=(r-q)S_u\,du+S_u\sqrt{Y_u}\,dW^S_u,\qquad S_0=s>0,\\
 dY_u&=\kappa(\theta-Y_u)\,du+\sigma\sqrt{Y_u}\,dW_u,\qquad Y_0=y\ge0,\\
 d\langle W^S,W\rangle_u&=\rho\,du.
 \end{aligned}
\end{equation}
The variance is the square-root process of Cox, Ingersoll and Ross
\cite{CIR}, whose equation has a pathwise unique nonnegative strong
solution by the Yamada--Watanabe theorem \cite{YW}. We use this solution
in \eqref{sde}. When the variance can reach zero, its behavior there is
governed by this stochastic model, and no additional absorbing
condition is imposed. The inequality $2\kappa\theta\ge\sigma^2$, under
which zero is not attained by Feller's classification \cite{Feller}, is
not assumed.

For remaining maturity $\tau\ge0$, define
\begin{equation}\label{value}
 P(\tau,s,y)=\sup_{0\le\eta\le\tau}
 \E_{s,y}\bigl[e^{-r\eta}(K-S_\eta)^+\bigr],
\end{equation}
where the supremum is over stopping times. Working with all finite $\tau$
avoids an artificial boundary in calendar time. A contract with maturity $T$
has price $V(t,s,y)=P(T-t,s,y)$.

Write
\begin{equation}\label{sets}
 \begin{aligned}
 \D&=(0,\infty)\times(0,\infty)\times(0,\infty),\\
 \C&=\{(\tau,s,y)\in\D\,\,\text{with}\,\,P(\tau,s,y)>(K-s)^+\},\\
 \G&=\partial_{\D}\C.
 \end{aligned}
\end{equation}
Spatial derivatives refer to $(s,y)$, or to $(x,y)$ after the change
$x=\log s$. The space $W^{1,2}_{p,\loc}$ consists of functions whose first
time derivative and whose spatial derivatives through order two belong
locally to $L^p$. The notation $C^{1,1}_{\mathrm{space},\loc}\cap C^{0,1}_{\tau,\loc}$
means that the spatial gradient is locally Lipschitz in the spatial variables,
uniformly in time on compact sets, and that the function is locally Lipschitz
in time. A trace taken from the continuation side is marked by a superscript
$+$ or by $s=b+$. For a symmetric matrix $A$ and a Hessian $H$ we write
$\tr(AH)$ for the contraction $\sum_{ij}A_{ij}H_{ij}$.

Our first main result describes the exercise boundary and the regularity of
the value function.

\begin{theorem}\label{main}
Under \eqref{assumptions}, the value function \eqref{value} has the following
properties.
\begin{enumerate}[label=\textup{(\roman*)},leftmargin=*]
\item There is a function $b$ from $(0,\infty)^2$ to $(0,K)$ such that
\begin{equation}\label{threshold}
 \C=\{(\tau,s,y)\in\D\,\,\text{with}\,\,s>b(\tau,y)\},\qquad
 \G=\{(\tau,b(\tau,y),y)\,\,\text{with}\,\,\tau,y>0\}.
\end{equation}
The function $b$ belongs to $C^\infty((0,\infty)^2)$ and satisfies
\begin{equation}\label{strictboundary}
 b_\tau(\tau,y)<0,\qquad b_y(\tau,y)<0.
\end{equation}
\item The function $P$ belongs to
\begin{equation}\label{valuereg}
 C^1(\D)\cap W^{1,2}_{p,\loc}(\D)
 \cap C^{1,1}_{\mathrm{space},\loc}(\D)
 \cap C^{0,1}_{\tau,\loc}(\D),\qquad 1<p<\infty.
\end{equation}
It is $C^\infty$ in $\C$ and is smooth up to each point of $\G$ from $\C$.
On the boundary,
\begin{equation}\label{fit}
 P=K-b,\qquad P_s=-1,\qquad P_y=0,\qquad P_\tau=0.
\end{equation}
In $\C$, one has $P_\tau>0$ and $1+P_s>0$.
\item The gamma from the continuation side exists and obeys
\begin{equation}\label{gamma}
 P_{ss}(\tau,b+,y)=
 \frac{2(rK-qb)}{y\bigl(b^2-2\rho\sigma b b_y+\sigma^2b_y^2\bigr)}>0.
\end{equation}
Consequently $P$ is not twice continuously differentiable across $\G$.
\end{enumerate}
All assertions are local and stay away from $\tau=0$ and from $y=0$.
\end{theorem}

For the weighted statement, put
\begin{equation}\label{logdefs}
 U(\tau,x,y)=P(\tau,e^x,y),\qquad
 w(\tau,x,y)=U(\tau,x,y)-K+e^x,\qquad
 \beta(\tau,y)=\log b(\tau,y),
\end{equation}
and let
\begin{equation}\label{matrixM}
 M=\begin{pmatrix}1&\rho\sigma\\\rho\sigma&\sigma^2\end{pmatrix},
 \qquad \lambda_M=\lambda_{\min}(M)>0.
\end{equation}
The function $w$ is nonnegative everywhere. Above the strike we continue the
affine put branch $K-e^x$ purely as an algebraic device. The second main
result gives the Hessian of $w$ on the free boundary and a bound for it that
is uniform as the variance tends to zero.

\begin{theorem}\label{weightedmain}
Under \eqref{assumptions}, the continuation-side spatial Hessian of $w$ on
$x=\beta(\tau,y)$ is
\begin{equation}\label{hessianmain}
 D^2_{x,y}w^+=
 \frac{2(rK-qb)}{y\bigl(1-2\rho\sigma\beta_y+\sigma^2\beta_y^2\bigr)}
 \begin{pmatrix}1&-\beta_y\\-\beta_y&\beta_y^2\end{pmatrix}.
\end{equation}
In particular, at every $\tau,y>0$,
\begin{equation}\label{weightedbound}
 y\norm{D^2_{x,y}w^+}_{\mathrm{op}}\le\frac{2rK}{\lambda_M}.
\end{equation}
For every finite $Y_*>0$, the corresponding traces of the price satisfy
\begin{equation}\label{weightedU}
 \sup_{\tau>0,\ 0<y\le Y_*}
 y\norm{D^2_{x,y}U(\tau,\beta(\tau,y)+,y)}_{\mathrm{op}}
 \le\frac{2rK}{\lambda_M}+Y_*K.
\end{equation}
\end{theorem}

For $\tau>0$, define the actual zero-variance exercise threshold by
\[
 B(\tau)=\sup\{s\in(0,K):P(\tau,s,0)=K-s\}.
\]
The natural CIR process, rather than an absorbing variance boundary,
determines this value. The next result describes what can be proved
at the endpoint without assuming its temporal smoothness. We write
$b(\tau,0)=B(\tau)$, and for $s<B(\tau)$ close to $B(\tau)$ we let
$\Gamma(\tau,s)$ denote the variance at which the boundary passes
through the stock $s$, so that the free boundary near the axis is the
graph $y=\Gamma(\tau,s)$.

\begin{theorem}\label{mainaxis}
Under \eqref{assumptions}, the following statements hold.
\begin{enumerate}[label=\textup{(\roman*)},leftmargin=*]
\item The axis threshold satisfies $0<B(\tau)<K$ and is locally
Lipschitz in remaining maturity. More precisely, for every compact
$I\Subset(0,\infty)$ there is $C_I$ such that
\begin{equation}\label{eq:mainaxis-lipschitz}
 0\le b(\tau,y)-b(\tau+h,y)\le C_I(1+y)h,
 \qquad \tau\in I,\ y\ge0,\ 0<h<h_I,
\end{equation}
so that the boundary moves with bounded speed in maturity uniformly
down to zero variance. The boundary extends continuously to
$[0,\infty)\times[0,\infty)$ by $b(\tau,0)=B(\tau)$ and $b(0,y)=K$,
and $b(\tau,y)\to K$ as $\tau\searrow0$ uniformly on bounded
variance intervals.
\item On compact intervals of positive maturities,
\begin{gather*}
 0\le P(\tau+h,s,y)-P(\tau,s,y)\le C(1+y)h,\\
 0\le P(\tau,s,y)-K+s\le Cy
 \qquad(s\le B(\tau),\ y\text{ small}),
\end{gather*}
and there are $c,C,y_0>0$ such that
\begin{equation}\label{eq:mainaxis-premium}
 cy\le P(\tau,B(\tau),y)-K+B(\tau)\le Cy\quad(0<y<y_0)
\end{equation}
uniformly for $\tau$ in the compact interval.
\item For every $\tau>0$,
\begin{equation}\label{eq:mainaxis-separation}
 \frac{B(\tau)-b(\tau,y)}y\longrightarrow+\infty,
 \qquad b_y(\tau,y)\longrightarrow-\infty\quad(y\searrow0).
\end{equation}
The inverse spatial curve extends as a regular $C^1$ curve tangent
to the zero-variance axis at $(B(\tau),0)$, and both partial derivatives
of the inverse graph vanish along the slice,
$\Gamma_s(\tau,s)\to0$ and $\Gamma_\tau(\tau,s)\to0$ as
$s\nearrow B(\tau)$.
\item Put $\nu=2\kappa\theta/\sigma^2$. Locally uniformly in positive
remaining maturity, for sufficiently small $y$,
\begin{equation}\label{eq:mainaxis-gap-upper}
 0\le\log\frac{B(\tau)}{b(\tau,y)}\le C
 \begin{cases}
 y^{\nu/(\nu+1)},&0<\nu<1,\\
 \{y\log(e/y)\}^{1/2},&\nu=1,\\
 y^{1/2},&\nu>1.
 \end{cases}
\end{equation}
These are upper bounds, not asymptotic equivalents.
\item The weighted spatial Hessian from the continuation side has the exact limit
\begin{equation}\label{eq:mainaxis-weighted}
 \lim_{y\searrow0}yD^2_{x,y}w(\tau,\beta(\tau,y)+,y)
 =\frac{2(rK-qB(\tau))}{\sigma^2}
 \begin{pmatrix}0&0\\0&1\end{pmatrix}.
\end{equation}
\item The variance derivative does not extend continuously to the axis
contact point. It vanishes on the contact set, whereas
\begin{equation}\label{eq:mainaxis-py}
 \limsup_{y\searrow0}P_y(\tau,B(\tau),y)\ge c>0,
\end{equation}
with the constant of \eqref{eq:mainaxis-premium}. Smooth fit therefore
fails in the degenerate direction at the axis.
\end{enumerate}
The $C^1$ endpoint assertion concerns each spatial
curve. It is not a claim of $C^1$ regularity in time at the axis, of
differentiability of $B$, or of continuity of the stock derivative
$P_s$ at the axis contact point. Remarks~\ref{rem:axis-smooth-fit}
and~\ref{rem:axis-C1-surface} explain what these statements would
require.
\end{theorem}

The bounds \eqref{weightedbound} and \eqref{weightedU} concern the traces on
the free boundary and are not weighted $C^{1,1}$ estimates on neighborhoods
of $y=0$. Section~\ref{sec:local} collects the properties of the value
function, proves its local Sobolev regularity and reduces the local obstacle
problem to the setting of \cite{AK}. Section~\ref{sec:geometry} proves
Theorems~\ref{main} and \ref{weightedmain} and derives further consequences.
Section~\ref{sec:axis} proves Theorem~\ref{mainaxis}, derives exact
vertical identities, including the additional maturity term, and
records the questions that remain open at the axis.
Section~\ref{sec:numerics} computes the boundary surface numerically and
compares the computed boundary gamma with \eqref{gamma}.

\section{The local obstacle problem}\label{sec:local}

In log coordinates the generator and the discounted parabolic operator read
\begin{equation}\label{operators}
 \begin{aligned}
 \Lh u&=\tr\bigl(\Sigma(y)D^2u\bigr)+m(y)\cdot Du,\qquad \Sigma(y)=\frac y2M,\\
 m(y)&=\bigl(r-q-y/2,\ \kappa(\theta-y)\bigr),\\
 \op u&=u_\tau-\Lh u+ru.
 \end{aligned}
\end{equation}
On every compact subset of $\R\times(0,\infty)$ the matrix $\Sigma$ is uniformly
positive definite. We first collect the properties of the value function that
follow from its probabilistic definition.

\begin{proposition}\label{probprops}
The value function is continuous, satisfies $0\le P\le K$, is nondecreasing
in $\tau$ and in $y$, and is nonincreasing and convex in $s$. It is
$1$-Lipschitz in $s$ and extends continuously to $P(\tau,0,y)=K$. For every
$\tau>0$ and $y>0$ it is strictly positive at every $s>0$. There is a
threshold $b(\tau,y)\in(0,K)$ such that the contact set at $(\tau,y)$ is the
interval $(0,b(\tau,y)]$. The threshold is nonincreasing in each of $\tau$
and $y$.
\end{proposition}

\begin{proof}
Continuity and variance monotonicity are Proposition~4.1 of \cite{LTprop},
and positivity of the threshold is Proposition~4.5 of the same paper. Both
results allow the Feller condition to fail. We include the remaining
elementary arguments to fix signs and coordinates.

For a fixed initial variance the stock has the form $S_u=sZ_u$, where $Z$
does not depend on the initial stock price. For each stopping time the
expected payoff is convex and nonincreasing in $s$, and a supremum preserves
both properties. The process $e^{-ru}Z_u$ is a nonnegative supermartingale
because $q\ge0$. Optional sampling at bounded stopping times therefore gives
\[
 \abs{\E[e^{-r\eta}(K-s_1Z_\eta)^+]
       -\E[e^{-r\eta}(K-s_2Z_\eta)^+]}
 \le\abs{s_1-s_2}.
\]
The same bound holds for the supremum. Immediate exercise and boundedness
then give $P(\tau,s,y)\to K$ as $s\searrow0$. Monotonicity in remaining
time follows by enlarging the admissible set of stopping times.

For positivity, fix a positive deterministic time $h\le\tau$. Write
$W^S=\rho W+\sqrt{1-\rho^2}\,W^\perp$ with $W^\perp$ independent of $W$.
Conditionally on $W$, the term $\sqrt{1-\rho^2}\int_0^h\sqrt{Y_u}\,dW^\perp_u$
in $\log S_h$ is Gaussian with variance $(1-\rho^2)\int_0^hY_u\,du$, which
is positive almost surely because $Y$ is continuous and $Y_0=y>0$. Hence
$\PP(S_h<K)>0$, and the discounted European payoff at time $h$ has strictly
positive expectation. This proves $P>0$ and places $s=K$ in the
continuation region.

Finally, $s\mapsto P(\tau,s,y)-(K-s)$ is nonnegative and nondecreasing by the
$1$-Lipschitz property. Its zero set is an initial interval, closed at its
right endpoint by continuity. The cited threshold result says that this
endpoint is positive. Positivity of $P$ at $K$ gives $b<K$. Increasing
either $\tau$ or $y$ can only shrink the zero set, which gives the asserted
monotonicities.
\end{proof}

For later use, the same argument together with differentiability in $s$
gives the stronger inequality
\begin{equation}\label{vxpositive}
 s(1+P_s(\tau,s,y))\ge P(\tau,s,y)-K+s.
\end{equation}
Indeed, the supporting line inequality for the convex function $P$ at $s$
gives $K=P(\tau,0,y)\ge P(\tau,s,y)-sP_s(\tau,s,y)$.

In log coordinates, Proposition~\ref{probprops} says that $w$ is
nondecreasing separately in $\tau$, $x$ and $y$. Continuity of the threshold
is not used in what follows.

A direct computation gives
\begin{equation}\label{forcing}
 \op(K-e^x)=rK-qe^x=:f(x).
\end{equation}
Near every contact point with $\tau,y>0$ one has $x<\log K$ and
\begin{equation}\label{fpositive}
 f(x)>0.
\end{equation}
This follows from $q\le r$ and $e^x<K$, and it includes the case $q=r$.

The dynamic programming principle for \eqref{value}, stopped before leaving
a bounded cylinder in which $y$ stays away from zero, gives the local
viscosity inequality
\begin{equation}\label{VI}
 \min\{w,\op w+f\}=0.
\end{equation}
Only cylinders with $x<\log K$ are used for \eqref{VI}. One can obtain it
directly by applying It\^o's formula to a smooth test function up to the
minimum of a short time, the exit time and an admissible exercise time. At a
point with positive gap, exercise is locally suboptimal, so both viscosity
inequalities reduce to the linear equation. At a point with zero gap, waiting
for a short stopped interval gives the supersolution inequality, while the
subsolution inequality is automatic from the zero obstacle. The reversal from
calendar time to $\tau$ produces the sign $\partial_\tau-\Lh+r$ in
\eqref{VI}. These are local arguments for the continuous value function, and
no boundary condition at $y=0$ enters. Dynamic programming and the viscosity
identification for optimal stopping are discussed in \cite{PS}, the
general theory of viscosity solutions is in \cite{CIL}, and the
Heston formulation appears in \cite{LTvi,NC}. The next lemma gives the local
Sobolev regularity of solutions of \eqref{VI} away from $y=0$.

\begin{lemma}\label{penalty}
Let $Q$ be a bounded parabolic cylinder whose closure lies in
$\{\tau>0,y>0,x<\log K\}$, and suppose $f\ge f_0>0$ there. Every continuous
viscosity solution $w\ge0$ of \eqref{VI} belongs to $W^{1,2}_{p,\loc}(Q)$
for every $1<p<\infty$. Its spatial gradient is locally continuous, and
\begin{equation}\label{ae}
 \op w=-f\chi_{\{w>0\}}\qquad\text{almost everywhere in }Q.
\end{equation}
Moreover $w\in C^{1,\alpha}_p$ locally for every $0<\alpha<1$, in the
parabolic H\"older notation of \cite{AK}.
\end{lemma}

\begin{proof}
Shrink $Q$ slightly if necessary and regard the actual continuous trace
$h=w|_{\partial_p Q}\ge0$ as prescribed data. On this one fixed cylinder solve
\begin{equation}\label{penalized}
 \op w_\varepsilon=-f+\varepsilon^{-1}(w_\varepsilon)^-,
 \qquad w_\varepsilon=h\quad\text{on }\partial_pQ,
\end{equation}
where $z^-=\max\{-z,0\}$. Once the nonlinearity is moved to the left-hand
side it is nondecreasing in the unknown. Existence and comparison follow
from the standard semilinear parabolic Dirichlet theory, after a smooth
approximation of the boundary data if needed. Only interior estimates will
be used.

Set $\bar f=\norm{f}_{L^\infty(Q)}$ and $\bar h=\max_{\partial_pQ}h$. The constants
$-\varepsilon \bar f$ and $\bar h$ are respectively a subsolution and a supersolution
of \eqref{penalized}. Indeed,
\[
 \op(-\varepsilon \bar f)=-r\varepsilon \bar f\le -f+\bar f,
 \qquad \op \bar h=r\bar h\ge-f.
\]
Comparison yields
\begin{equation}\label{penaltybounds}
 -\varepsilon \bar f\le w_\varepsilon\le \bar h,
 \qquad 0\le\varepsilon^{-1}(w_\varepsilon)^-\le \bar f.
\end{equation}
In particular the right-hand side of \eqref{penalized} is bounded uniformly
in $\varepsilon$. For $Q'\Subset Q$, the interior $L^p$ estimates for linear
parabolic equations give
\begin{equation}\label{wpestimate}
 \norm{w_\varepsilon}_{W^{1,2}_p(Q')}
 \le C_{p,Q',Q}\bigl(\bar h+\bar f+1\bigr).
\end{equation}
Here and below all such constants may depend on the distance from $y=0$.
The interior estimates, the Sobolev embeddings and the comparison principles
used are the standard ones for smooth uniformly parabolic operators, see
\cite{Lieberman,LSU}.

The bounds \eqref{penaltybounds} and the bounded
right-hand side provide a common modulus of continuity near $\partial_pQ$
through the usual Dirichlet barriers on a smooth cylinder. Together with
\eqref{wpestimate} this gives local compactness and attainment of the same
trace $h$. A subsequential limit $\bar w$ is nonnegative. On $\{\bar w>0\}$
the penalty vanishes locally and $\op\bar w=-f$. Everywhere one has
$\op w_\varepsilon+f\ge0$, and viscosity stability gives
$\op\bar w+f\ge0$. Thus $\bar w$ solves \eqref{VI} with trace $h$.
Comparison for this proper uniformly parabolic obstacle problem identifies
$\bar w=w$. Weak compactness in \eqref{wpestimate} then proves the asserted
Sobolev regularity.

The Sobolev property of zero sets gives $w_\tau=0$ and $Dw=0$ almost everywhere
on $\{w=0\}$. Applying the same property to each spatial derivative, using
spatial slices and Fubini, gives $D^2w=0$ almost everywhere there. Therefore
$\op w=0$ almost everywhere on the contact set. Combined with the equation
on $\{w>0\}$, this is \eqref{ae}.

In two spatial dimensions, choose $p>4/(1-\alpha)$. The parabolic Sobolev
embedding gives local $C^{1,\alpha}_p$ regularity. Since $p$ is arbitrary,
this holds for every $\alpha<1$. In particular $Dw$ is continuous, and as
$w\ge0$ its spatial gradient vanishes at every contact point.

The argument uses only the uniform parabolicity of $\op$ on $Q$, the
bounds of its coefficients there, and the sign of its zeroth-order
coefficient. It therefore applies verbatim to any smooth uniformly
parabolic operator with a bounded nonnegative zeroth-order coefficient,
and the constants in \eqref{wpestimate} depend only on the ellipticity
and coefficient bounds, on $\bar f$, $\bar h$ and $p$. This uniformity
is used in Section~\ref{sec:axis} for rescaled operators whose discount
tends to zero.
\end{proof}

Lemma~\ref{penalty} applies near every free-boundary point, and the remaining
points of $\D$ are handled directly.

\begin{remark}\label{strike}
The kink of the put payoff causes no gap in the preceding localization.
Every free-boundary point satisfies $s<K$, and every point with $s\ge K$
lies in the open continuation region. In that region $P$ solves the
homogeneous linear parabolic equation with smooth coefficients and is smooth
by interior regularity. Points in the interior of the contact set have
$P=K-s$. Applying Lemma~\ref{penalty} near the remaining points establishes
local $W^{1,2}_p$ regularity everywhere in $\D$.
\end{remark}

The regularity theory of \cite{AK} concerns equations without first-order
and zeroth-order terms. We now show that the drift and the discount can be
absorbed into the principal part by adding one spatial variable. The
reduction works in any spatial dimension. Let $O\subset\R^d$ be bounded and
consider
\begin{equation}\label{generalP}
 \mathscr Q u=u_t-\tr\bigl(A(z)D_z^2u\bigr)-B(z)\cdot D_zu+c(z)u,
\end{equation}
where the coefficients are smooth on a neighborhood of $\overline O$,
$A$ is symmetric and uniformly positive definite, and
\begin{equation}\label{discountpositive}
 c(z)\ge c_*>0.
\end{equation}
The following lemma absorbs the lower-order terms of $\mathscr Q$ into the
principal part.

\begin{lemma}\label{liftlemma}
There are an interval $I$ containing zero, a positive smooth function $\phi$
on $I$ with $\phi(0)=1$, and a smooth symmetric uniformly positive definite
matrix $\widetilde A(z,\zeta)$ on $O\times I$ such that, for
$\widetilde u(t,z,\zeta)=\phi(\zeta)u(t,z)$,
\begin{equation}\label{liftidentity}
 \partial_t\widetilde u-
 \tr\bigl(\widetilde A\,D^2_{z,\zeta}\widetilde u\bigr)
 =\phi(\zeta)\mathscr Q u.
\end{equation}
The identity holds classically for smooth $u$ and almost everywhere for
$u\in W^{1,2}_{p,\loc}$.
\end{lemma}

\begin{proof}
Choose $\zeta_*\in(0,\pi/2)$ and a small interval $I$ around zero, both to
be fixed below, and put
\begin{equation}\label{phidef}
 \phi(\zeta)=\frac{\cos(\zeta+\zeta_*)}{\cos\zeta_*},\qquad
 \ell(z,\zeta)=\frac{\phi(\zeta)}{2\phi'(\zeta)}B(z)
             =-\frac{B(z)}{2\tan(\zeta+\zeta_*)}.
\end{equation}
Take $I$ so small that $0<\zeta+\zeta_*<\pi/2$ on $I$. Then $\phi>0$,
$\phi'\ne0$ and $\phi''=-\phi$. Define
\begin{equation}\label{liftmatrix}
 \widetilde A(z,\zeta)=
 \begin{pmatrix}A(z)&\ell(z,\zeta)\\\ell(z,\zeta)^T&c(z)\end{pmatrix}.
\end{equation}
Let $L=\sup_{\overline O} B^TA^{-1}B<\infty$. Choose $\zeta_*$ close enough
to $\pi/2$, and then $I$ small enough, that
\[
 \frac{L}{4\tan^2(\zeta+\zeta_*)}\le\frac{c_*}{2}
 \quad\text{on }I.
\]
The Schur complement then satisfies
\begin{equation}\label{schur}
 c-\ell^TA^{-1}\ell\ge c_*/2.
\end{equation}
To see uniform ellipticity directly, for $(\xi,\eta)\in\R^d\times\R$ write
\[
 (\xi,\eta)^T\widetilde A(\xi,\eta)
 = (\xi+A^{-1}\ell\eta)^TA(\xi+A^{-1}\ell\eta)
   +(c-\ell^TA^{-1}\ell)\eta^2.
\]
Uniform bounds on $A$ and on $A^{-1}\ell$ together with \eqref{schur} give
positive uniform lower and upper bounds for the quadratic form. Shrinking to
a relatively compact subinterval of $I$ if necessary keeps all coefficients
and their derivatives bounded.

Finally,
\[
 D_z^2\widetilde u=\phi D_z^2u,\quad
 D_z\partial_\zeta\widetilde u=\phi'D_zu,\quad
 \partial_{\zeta\zeta}\widetilde u=-\phi u.
\]
Consequently
\[
 \tr\bigl(\widetilde A\,D^2\widetilde u\bigr)
 =\phi\tr\bigl(AD^2u\bigr)+2\phi'\ell\cdot Du-c\phi u
 =\phi\bigl(\tr(AD^2u)+B\cdot Du-cu\bigr),
\]
which proves \eqref{liftidentity}. Multiplication by a fixed smooth function
preserves the required local Sobolev derivatives, which proves the last
assertion.
\end{proof}

The construction depends on the ellipticity of $A$ and on the positive
discount.

\begin{remark}\label{liftlocal}
The lifting is local and uses \eqref{discountpositive}. It is not a uniform
embedding when the ellipticity of $A$ degenerates. For the Heston problem,
$A=\Sigma(y)$ and $c=r$, and the interval and the ellipticity constants may depend
on a positive lower bound for $y$.
\end{remark}

We state precisely the consequence of \cite{AK} that is used. Let
$\mathcal A(Z)$ be a smooth, symmetric and uniformly positive definite
matrix, let $h$ be smooth with $h\le-h_0<0$, and let $v\ge0$ with $v_t\ge0$
solve
\begin{equation}\label{AKequation}
 v_t-\tr\bigl(\mathcal A(Z)D^2v\bigr)=h(Z)\chi_{\{v>0\}}
\end{equation}
almost everywhere in a cylinder. The operator $F(N,Z)=\tr(\mathcal A(Z)N)$
is smooth, linear and hence convex in $N$, uniformly elliptic, and satisfies
$F(0,Z)=0$, as required in \cite{AK}. Assume also that locally
$v\in W^{1,2}_p$ for every finite $p>1$ and $v\in C_p^{1,\alpha}$ for every
$\alpha<1$. These are the local solution requirements of Definition~2.1 in
\cite{AK}, and the local boundedness requirements on $v$ and $h$ hold
automatically here.

Theorem~5.1 of \cite{AK} gives local boundedness of $D^2v$ and of $v_t$.
Theorem~1.1 of \cite{AK} gives the following alternative at every
free-boundary point $(t_0,Z_0)$ for the parabolic blow-ups
\begin{equation}\label{blowup}
 v_\varepsilon(t,Z)=\varepsilon^{-2}
 v(t_0+\varepsilon^2t,Z_0+\varepsilon Z).
\end{equation}
Either the point is regular and $v_\varepsilon$ converges to
$\gamma_0(e\cdot Z)_+^2$ with $\gamma_0>0$ and $\abs e=1$, or along a subsequence of every
sequence of scales the blow-ups converge to a time-independent quadratic
$Z^T\mathcal HZ$ with $\mathcal H\ge0$. The limit is nonzero by the nondegeneracy Lemma~5.4 of
\cite{AK}. The convergence takes place in $C^{1,\alpha}_x\cap C^\alpha_t$ on
compact sets and therefore includes local uniform convergence at $t=0$. The
regular points form a relatively open subset of the free boundary which is
locally a $C^\infty$ space-time hypersurface with nonzero spatial normal,
and by Proposition~8.5 of \cite{AK} the spatial Hessian has a continuous
trace from the continuation side there. The covering theorem for the singular set in \cite{AK},
which requires $F$ to be independent of $Z$, is not used. Theorem~1.1 of
\cite{AK} is stated for operators that depend on the spatial variable, but
two steps of its proof are written for operators that do not, and the
following remark explains why they extend to \eqref{AKequation}.

\begin{remark}\label{AKremark}
The convexity of blow-ups is derived in Section~7 of \cite{AK} from the
semiconvexity estimate of Proposition~6.1 there, proved when $F$ does not
depend on $Z$. A blow-up $v_0$ at $(t_0,Z_0)$ does not depend on time by the
first estimate in that proposition, which holds for every $F$ considered
there, it solves the obstacle problem with the frozen operator
$\tr(\mathcal A(Z_0)D^2v_0)$ and the constant source $h(Z_0)$, and it grows
at most quadratically by Lemma~5.3 of \cite{AK}. The semiconvexity estimate
therefore applies to the rescalings $R^{-2}v_0(RZ)$ at $Z=0$, and letting
$R\to\infty$ shows that $v_0$ is convex. The Lipschitz bound on $F$ in $Z$
assumed in Section~8 of \cite{AK} is used only for matrices bounded by the
optimal regularity estimate, on which
$\nabla_ZF(N,Z)=\tr(\nabla_Z\mathcal A(Z)N)$ is bounded because $\mathcal A$
is smooth.
\end{remark}

Combining the lifting with this alternative gives a criterion for regularity
in terms of the shape of the contact set.

\begin{proposition}\label{contactcriterion}
Suppose that the coefficients of \eqref{generalP} satisfy the hypotheses of
Lemma~\ref{liftlemma}, and that $f\ge f_0>0$ is smooth and independent of
time. Suppose $u\ge0$, $u_t\ge0$, $u\in W^{1,2}_{p,\loc}\cap C_p^{1,\alpha}$
for every finite $p>1$ and every $\alpha<1$, and
\begin{equation}\label{generalobstacle}
 \mathscr Q u=-f(z)\chi_{\{u>0\}}.
\end{equation}
Then $u$ has locally bounded spatial second derivatives and time derivative.
Let $(t_0,z_0)$ be a free-boundary point. If there are a nonempty open cone
$\mathcal K\subset\R^d$ and $\delta>0$ such that
\begin{equation}\label{conecondition}
 u(t_0,z_0+z)=0\qquad (z\in\mathcal K\cap B_\delta),
\end{equation}
then the free boundary is a smooth space-time hypersurface with nonzero
spatial normal near $(t_0,z_0)$, and $u$ is smooth up to this boundary from
its positive side.
\end{proposition}

\begin{proof}
Apply Lemma~\ref{liftlemma} and set $Z=(z,\zeta)$. Since $\phi>0$,
\begin{equation}\label{product}
 \{\widetilde u>0\}=\{u>0\}\times I,
 \qquad \partial\{\widetilde u>0\}=\partial\{u>0\}\times I
\end{equation}
locally. The lifted equation is
\[
 \widetilde u_t-\tr\bigl(\widetilde A(Z)D^2\widetilde u\bigr)
 =-\phi(\zeta)f(z)\chi_{\{\widetilde u>0\}}.
\]
This is \eqref{AKequation} with $\mathcal A=\widetilde A$ and
$h(Z)=-\phi(\zeta)f(z)$, and $\widetilde A$ is uniformly positive definite by
\eqref{schur}. The source is smooth and is bounded strictly below zero on a
smaller neighborhood. The inequality $\widetilde u_t\ge0$ follows from
$u_t\ge0$. Multiplication by $\phi$ and the introduction of a passive
variable preserve every local Sobolev and H\"older assumption. Thus each
hypothesis listed for \eqref{AKequation} is satisfied. The optimal local
derivative bounds follow from Theorem~5.1 of \cite{AK} after restriction to
$\zeta=0$.

At $(t_0,z_0,0)$, condition \eqref{conecondition} and the product identity
show that every blow-up vanishes at time zero on $\mathcal K\times\R$.
Indeed, each fixed $(z,\zeta)$ in this set belongs to the rescaled contact
set at all sufficiently small scales. A nonzero quadratic polynomial cannot
vanish on a nonempty open subset of $\R^{d+1}$, so the singular alternative
is impossible. The point is regular, and the same holds along the fibre in
the added variable.

The lifted free boundary is therefore smooth. Its product structure makes the
$\zeta$ direction tangent, so the spatial normal has zero $\zeta$ component
and a nonzero $z$ component. Hence its intersection with $\zeta=0$ is
transverse and is a smooth hypersurface with nonzero spatial normal, which is
the assertion for $u$.

It remains to justify the one-sided smoothness stated in the proposition.
The lifted Hessian has a continuous trace from the positive side by Proposition~8.5
of \cite{AK}, and hence so does $D_z^2u$. After a spatial rotation the smooth
boundary can be written as a smooth moving graph, which we flatten. In its
positive half-cylinder $u$ solves a linear uniformly parabolic equation with
smooth coefficients and smooth right-hand side, with homogeneous Dirichlet
data on the flat lateral boundary. Local boundary Schauder estimates and
their higher-order versions give $C^\infty$ regularity up to the smaller
lateral boundary. Since the localization stays away from the bottom of the
cylinder, no compatibility condition is needed, see \cite{Lieberman}.
\end{proof}

Monotonicity in every spatial coordinate provides such a cone at every
free-boundary point.

\begin{corollary}\label{orthant}
Under the hypotheses of Proposition~\ref{contactcriterion}, if $u$ is
nondecreasing in each of its $d$ spatial coordinates, then every
free-boundary point is regular and the whole interior free boundary is
locally smooth.
\end{corollary}

\begin{proof}
At a contact point $(t_0,z_0)$, nonnegativity and monotonicity give
$0\le u(t_0,z_0+z)\le u(t_0,z_0)=0$ whenever every coordinate of $z$ is
negative and $z$ is small. The negative orthant is an open cone, so
Proposition~\ref{contactcriterion} applies.
\end{proof}

\section{Proof of the main results}\label{sec:geometry}

We now apply the criterion and supply the graph and monotonicity arguments
that are specific to the Heston problem. All normal vectors in this section
are spatial unit normals directed into the continuation region. The first
step shows that every interior free-boundary point is regular.

\begin{proposition}\label{smoothsurface}
In log coordinates the interior free boundary is a smooth space-time
hypersurface with nonzero spatial normal. The gap $w$ is smooth up to it from
the continuation side. Moreover $w$ has locally bounded spatial second
derivatives and time derivative.
\end{proposition}

\begin{proof}
Fix a boundary point. Proposition~\ref{probprops} places it below the strike
and at positive variance and positive remaining maturity. Choose a
surrounding cylinder on which $f\ge f_0>0$ and $\Sigma(y)$ is uniformly elliptic.
Lemma~\ref{penalty} gives the required local solution class and the
almost-everywhere equation. The discount is $r>0$. The three monotonicities
$w_\tau,w_x,w_y\ge0$ follow from Proposition~\ref{probprops} in the weak
sense, and then pointwise wherever the derivatives are continuous.
Corollary~\ref{orthant} applies with two spatial variables, $A=\Sigma$, $B=m$ and $c=r$.

Away from the free boundary the value is either the smooth payoff branch or a
solution of a smooth linear uniformly parabolic equation. A finite local
cover therefore yields the asserted bounded derivatives on every compact
subset of the interior domain.
\end{proof}

Let $n=(n_x,n_y)$ be the inward spatial unit normal at a boundary point
in log coordinates. Spatial smooth fit follows already from
Lemma~\ref{penalty}, since $w=0$ and $Dw=0$ on the contact set. The
one-sided smoothness from Proposition~\ref{smoothsurface} gives the following
traces.

\begin{lemma}\label{rankone}
At every interior free-boundary point,
\begin{equation}\label{rankonetraces}
 w=w_x=w_y=w_\tau=0,\qquad
 D^2w^+=\gamma\,n\otimes n,\qquad
 \gamma=\frac{f(x)}{n^T\Sigma(y)n}>0.
\end{equation}
All these traces are continuous along the boundary.
\end{lemma}

\begin{proof}
Since the smooth hypersurface has a nonzero spatial normal, it has a local
parameterization $(\tau,Z(\tau,\xi))$ in which time is a free parameter.
Differentiating $w(\tau,Z(\tau,\xi))=0$ in $\tau$ and using that the spatial
gradient vanishes on the boundary gives $w_\tau=0$ there. Differentiating
$Dw=0$ in a spatial tangential direction shows that the symmetric spatial
Hessian annihilates the tangent line, so it has the form $\gamma n\otimes n$.
Taking the trace of $\op w=-f$ from the continuation side gives
$\tr(\Sigma D^2w^+)=f$, and hence the formula for $\gamma$. Positivity follows
from $f>0$ and ellipticity. The regularity already proved gives continuity
of every trace used here.
\end{proof}

Smoothness alone does not show that the free boundary is a graph over the
stock direction. This follows from a Hopf argument for $w_x$.

\begin{lemma}\label{stocktransverse}
At every interior boundary point $n_x>0$. Consequently there is a smooth
local graph $x=\beta(\tau,y)$, and these local graphs combine to
$\beta\in C^\infty((0,\infty)^2)$.
\end{lemma}

\begin{proof}
In the continuation region put $v=w_x$. Since the coefficients in
\eqref{operators} do not depend on $x$, differentiation gives
\begin{equation}\label{veq}
 \op v=qe^x\ge0.
\end{equation}
By \eqref{vxpositive},
\begin{equation}\label{vstrict}
 v=e^x(1+P_s)\ge P-K+e^x=w>0\quad\text{in the continuation region}.
\end{equation}
The boundary trace of $v$ is zero. All coefficients are smooth and
uniformly parabolic near the point under consideration. The boundary is a
smooth moving hypersurface with nonzero spatial normal, and $v$ is smooth
from inside. The parabolic Hopf boundary lemma gives
\begin{equation}\label{hopfv}
 \partial_n v>0.
\end{equation}
One way to apply the lemma is to flatten the smooth moving boundary by a
time-dependent spatial change of variables. Its time dependence adds only a
bounded drift, and a small backward half-cylinder lies on the continuation
side. The nonnegative supersolution \eqref{veq} with zero lateral trace then
satisfies the Hopf inequality at a flat boundary, and the zeroth-order coefficient
$r>0$ causes no difficulty because the boundary minimum is zero.

On the other hand, \eqref{rankonetraces} gives
\[
 \partial_n v=n^TD^2w\,e_x=\gamma n_x.
\]
Since $\gamma>0$, \eqref{hopfv} implies $n_x>0$. The implicit function
theorem now gives a smooth graph in $x$. The unique threshold of
Proposition~\ref{probprops} identifies this local graph with
$\log b(\tau,y)$.

The same reasoning shows that the threshold has no jumps. Near the fixed
graph point, a short vertical segment in $x$ meets the smooth boundary
exactly once, with contact below and continuation above. For nearby
$(\tau,y)$ the same separating graph and the threshold property force that
crossing to be the unique threshold.
\end{proof}

A similar argument for $w_y$ gives strict monotonicity of the boundary in the
variance.

\begin{lemma}\label{variance}
The boundary satisfies $\beta_y<0$, and hence $b_y<0$, at every $\tau,y>0$.
\end{lemma}

\begin{proof}
Put $h=w_y\ge0$ in the continuation region. Differentiating
\eqref{operators} and using that $f$ does not depend on $y$ gives
\begin{equation}\label{commutator}
 \op h=\frac12\tr\bigl(MD^2w\bigr)-\frac12w_x-\kappa w_y.
\end{equation}
The equation for $w$ can be written as
\[
 \frac y2\tr\bigl(MD^2w\bigr)
 =w_\tau-(r-q-y/2)w_x-\kappa(\theta-y)w_y+rw+f.
\]
Substituting into \eqref{commutator} yields the exact identity
\begin{equation}\label{heq}
 \left(\op+\frac{\kappa\theta}{y}\right)h
 =\frac{w_\tau-(r-q)w_x+rw+f}{y}.
\end{equation}
At the boundary the numerator has trace $f>0$ by Lemma~\ref{rankone}. By
one-sided continuity, the right-hand side is strictly positive in a
sufficiently small continuation collar of the given point. In that collar
$h\ge0$ and $h$ has zero boundary trace.

In fact $h>0$ inside the collar. At an interior zero, nonnegativity would
give $h_\tau=0$, $Dh=0$ and $D^2h\ge0$, which contradicts the strictly
positive right-hand side of \eqref{heq}. The local Hopf lemma for
\eqref{heq} therefore gives $\partial_n h>0$ at the boundary. The
zeroth-order coefficient $r+\kappa\theta/y$ is bounded and positive on the
chosen collar, which is harmless since the boundary minimum is zero. From
\eqref{rankonetraces},
\[
 \partial_n h=\gamma n_y>0.
\]
Thus $n_y>0$. Since the inward unit normal to $x=\beta(\tau,y)$ is
\begin{equation}\label{normal}
 n=\frac{(1,-\beta_y)}{\sqrt{1+\beta_y^2}},
\end{equation}
we conclude that $\beta_y<0$.
\end{proof}

For the remaining maturity, positivity of $w_\tau$ has to be propagated from
the region above the strike.

\begin{lemma}\label{time}
One has $w_\tau>0$ throughout the continuation region and $\beta_\tau<0$ on
the boundary. Consequently $P_\tau>0$ in $\C$ and $b_\tau<0$.
\end{lemma}

\begin{proof}
Set $z=w_\tau$. Time monotonicity gives $z\ge0$, while time independence of
the coefficients and of $f$ gives
\begin{equation}\label{zeq}
 \op z=0\quad\text{in the continuation region}.
\end{equation}
We verify that $z$ cannot vanish at an interior continuation point, and this
includes the case $q=0$.

Fix $(\tau_0,x_0,y_0)$ in the continuation region and choose
$x_* >\max\{x_0,\log K\}$. The segment
$\{(\tau_0,x,y_0)\,\,\text{with}\,\,x_0\le x\le x_*\}$ is a compact subset
of the continuation region, since $w$ is nondecreasing in $x$ and positive at
the left endpoint. Openness gives a short past time strip, together with a
small spatial neighborhood of this segment, entirely inside the
continuation region.

At $(x_*,y_0)$ one has $P(0,e^{x_*},y_0)=0$, whereas $P(a,e^{x_*},y_0)>0$
for every $a>0$. Continuity at zero and smoothness at positive times imply
that for each sufficiently small $a>0$ there is $a_0\in(0,a)$ with
$P_\tau(a_0,e^{x_*},y_0)>0$. To justify this without differentiability at
zero, choose $0<\delta<a$ with $P(\delta,e^{x_*},y_0)<P(a,e^{x_*},y_0)$ and
apply the mean value theorem on $[\delta,a]$.

The vertical line at $(x_*,y_0)$ lies in the continuation region for all
positive times. Choose $a$ below the past time strip just described. A path
that is nondecreasing in time first follows that vertical line and then moves
along the spatial segment inside the strip to $(\tau_0,x_0,y_0)$. A finite
chain of overlapping interior parabolic cylinders along the path propagates
positivity of the nonnegative solution of \eqref{zeq} by the strong maximum
principle. Hence $z(\tau_0,x_0,y_0)>0$.

The boundary trace of $z$ is zero by Lemma~\ref{rankone}. The local Hopf
lemma now gives $\partial_n z>0$. Differentiate the identity
$w_x(\tau,\beta(\tau,y),y)=0$ along the boundary in $\tau$ to get
\begin{equation}\label{timefitdiff}
 w_{x\tau}^++w_{xx}^+\beta_\tau=0.
\end{equation}
Since $z=0$ along each spatial boundary slice, $Dz$ is normal to that slice,
so $z_x=n_x\partial_n z>0$. Also $w_{xx}^+=\gamma n_x^2>0$. Equation
\eqref{timefitdiff} therefore gives
\[
 \beta_\tau=-\frac{n_x\partial_n z}{w_{xx}^+}<0.
\]
\end{proof}

Differentiating the zero traces of $w_x$ and $w_y$ along $x=\beta(\tau,y)$
at a fixed time yields
\begin{equation}\label{tangenthessian}
 w_{xy}^+=-\beta_yw_{xx}^+,
 \qquad w_{yy}^+=\beta_y^2w_{xx}^+.
\end{equation}
Taking the trace of the equation for $w$ at the free boundary, all
lower-order terms and the time derivative vanish. Therefore
\begin{equation}\label{wxx}
 \frac y2\bigl(1-2\rho\sigma\beta_y+\sigma^2\beta_y^2\bigr)
 w_{xx}^+=rK-qb.
\end{equation}
The factor in parentheses is strictly positive, since it equals
$(1,-\beta_y)M(1,-\beta_y)^T$.

\begin{proof}[Proof of Theorem~\ref{main}]
Proposition~\ref{probprops} provides the threshold.
Proposition~\ref{smoothsurface} and Lemma~\ref{stocktransverse} prove that
$b=e^\beta$ is smooth and identify the whole relative boundary with its
graph. Lemmas~\ref{variance} and \ref{time} prove \eqref{strictboundary}.

Local Sobolev regularity follows from Lemma~\ref{penalty} and
Remark~\ref{strike}. The bounded spatial second derivatives and time
derivative follow from Proposition~\ref{smoothsurface}. These bounds are
preserved by the smooth coordinate change $s=e^x$ on compact subsets with
$s>0$, and they give the stated local Lipschitz properties. Spatial first
derivatives are continuous by the Sobolev embedding. At a boundary point the
smooth one-sided time derivative is zero by Lemma~\ref{rankone}, and on the
contact side it is also zero. Thus the time derivative exists and is
continuous across the graph. More explicitly, the smooth gap on the
continuation side has zero value and zero first-order traces on the smooth graph, so its
extension by zero to the contact set is a $C^1$ function. This proves
$P\in C^1(\D)$ and the identities \eqref{fit}. Interior linear regularity and
the one-sided boundary argument prove the remaining smoothness assertions.

Finally,
\[
 U_x=sP_s,\qquad U_{xx}=sP_s+s^2P_{ss},\qquad
 w_{xx}=U_{xx}+s.
\]
At $s=b$, smooth fit gives $w_{xx}^+=b^2P_{ss}^+$, while $\beta_y=b_y/b$.
Substitution into \eqref{wxx} gives \eqref{gamma}. Its numerator is positive
and its denominator is positive. On the contact side $P_{ss}=0$, so this
positive trace precludes $C^2$ regularity across the free boundary.
\end{proof}

\begin{proof}[Proof of Theorem~\ref{weightedmain}]
Equations \eqref{tangenthessian} and \eqref{wxx} give \eqref{hessianmain}.
The matrix of rank one in that formula has operator norm $1+\beta_y^2$. Hence
\[
 y\norm{D^2w^+}_{\mathrm{op}}
 =\frac{2(rK-qb)(1+\beta_y^2)}
 {(1,-\beta_y)M(1,-\beta_y)^T}
 \le\frac{2(rK-qb)}{\lambda_M}\le\frac{2rK}{\lambda_M}.
\]
Since $U=w+K-e^x$, its Hessian differs from that of $w$ by
$-e^x e_x\otimes e_x$. On the boundary $e^x=b<K$, and the triangle
inequality gives \eqref{weightedU}.
\end{proof}

The one-sided smoothness also gives a quadratic separation of the price from
the payoff and expresses the boundary slopes through the Greeks.

\begin{corollary}\label{expansion}
Let $E\Subset(0,\infty)^2$ be a compact set of $(\tau,y)$ parameters. There
are positive constants $c_E,C_E,\delta_E$ such that, whenever $(\tau,y)\in E$
and $0\le s-b(\tau,y)\le\delta_E$,
\begin{equation}\label{separation}
 c_E(s-b)^2\le P(\tau,s,y)-(K-s)\le C_E(s-b)^2.
\end{equation}
More precisely, uniformly on $E$,
\begin{equation}\label{taylor}
 P(\tau,s,y)-(K-s)
 =\tfrac12G(\tau,y)(s-b)^2+O((s-b)^3),
\end{equation}
where $G=P_{ss}(\tau,b+,y)>0$ is given by \eqref{gamma}. Also,
\begin{equation}\label{ratios}
 b_y=-\lim_{s\searrow b}\frac{P_y}{1+P_s},
 \qquad
 b_\tau=-\lim_{s\searrow b}\frac{P_\tau}{1+P_s}.
\end{equation}
\end{corollary}

\begin{proof}
The graph over $E$ is compact and lies at positive distance from $s=0$,
$y=0$, $\tau=0$ and $s=K$. Smooth one-sided regularity and compactness give
a common small continuation collar and a uniform bound on the third stock
derivative. Taylor's theorem, value matching and smooth fit give
\eqref{taylor}. The smooth positive function $G$ has a positive minimum and
a finite maximum on $E$, which yields \eqref{separation} after reducing
$\delta_E$.

To derive the ratios without dividing two unspecified zero traces, set
$R=P-K+s$ and write, by the integral form of Taylor's theorem,
\[
 R(\tau,s,y)=(s-b)^2 A(\tau,s,y),\qquad
 A(\tau,b,y)=\tfrac12G(\tau,y)>0.
\]
The function $A$ is smooth up to the graph from above. Differentiating with
respect to $s$, $y$ and $\tau$ gives
\[
 \begin{aligned}
 R_s&=2(s-b)A+(s-b)^2A_s,\\
 R_y&=-2(s-b)b_yA+(s-b)^2A_y,\\
 R_\tau&=-2(s-b)b_\tau A+(s-b)^2A_\tau.
 \end{aligned}
\]
Since $R_s=1+P_s$, $R_y=P_y$ and $R_\tau=P_\tau$, the limits follow.
\end{proof}

The ratios \eqref{ratios} express the boundary slopes as continuation-side
limits of observable sensitivities. When $q>r$, the argument remains valid at
every free-boundary point where the source $f$ is positive.

\begin{corollary}\label{largedividend}
Suppose $r>0$ and $q\ge0$, with the other parameters as in
\eqref{assumptions}, but allow $q>r$. At any interior free-boundary point
$(\tau_0,s_0,y_0)$ satisfying
\begin{equation}\label{strictforcingpoint}
 rK-qs_0>0,
\end{equation}
the local smoothness, the graph transversality, the strict boundary
monotonicities, the smooth-fit identities and the formulas \eqref{gamma} and
\eqref{hessianmain} remain valid in a neighborhood of that point.
\end{corollary}

\begin{proof}
The probabilistic properties in Proposition~\ref{probprops} hold for every
$q\ge0$. Condition \eqref{strictforcingpoint} supplies a neighborhood on
which $f\ge f_0>0$. Every local step of the lifting, the contact orthant, the
transversality argument and the variance argument therefore applies without
change. For the time argument, positivity of $P_\tau$ in the continuation
region uses only time monotonicity, strict positivity of the European put,
smoothness of the continuation equation and connectivity by increasing the
stock price, and these also hold for $q>r$. The curvature formulas are
algebraic consequences of the resulting local regularity.
\end{proof}

At a point with $rK-qs_0=0$ the source in \eqref{AKequation} is not bounded
away from zero, and the corollary makes no assertion there. The restriction
$q\le r$ in Theorem~\ref{main} excludes such points. We close this section
with the limits of the method at zero variance and at expiry.

\begin{remark}\label{scope}
The bound \eqref{weightedbound} is uniform along the interior free boundary
as $y\searrow0$, because its proof uses only the boundary equation and the
positive definiteness of $M$. The other estimates are not uniform: in the
lifting the Schur complement depends on $m^T\Sigma^{-1}m$, which is unbounded
near zero variance, and the ellipticity constants in the obstacle theorem
degenerate in the same way. The present proof therefore gives neither a
smooth extension of $b$ to $y=0$ nor a bound for $yD^2U$ on full
neighborhoods of $y=0$. The weighted theory of \cite{DF} concerns the
elliptic obstacle problem, and the theories of \cite{FP,WZ} concern linear
equations, so they do not settle the degenerate free-boundary geometry,
which Section~\ref{sec:axis} treats by local time, stopping time
comparisons, and uniform interior rescalings. At $\tau=0$ the payoff has a
kink at the strike. Proposition~\ref{prop:expiry} shows that the boundary
converges to the strike there, as in the one-dimensional case \cite{CC},
while its shape near expiry is not studied. In calendar time the results
hold locally for $t<T$ and $y>0$, and $\partial_t b(T-t,y)>0$ follows from
\eqref{strictboundary}.
\end{remark}

\section{Zero variance at positive remaining maturity}\label{sec:axis}

The estimates in this section retain the natural CIR behavior at zero.
Unless stated otherwise, all constants are allowed to depend on a compact
interval of strictly positive remaining maturities. Throughout the
section we write
\begin{equation}\label{eq:axis-notation}
 a=\kappa\theta,\quad \varsigma=\frac{\sigma^2}2,\quad
 \nu=\frac a\varsigma=\frac{2\kappa\theta}{\sigma^2},\quad
 \mu=r-q,\quad g(s)=(K-s)^+,\quad R=P-K+s,
\end{equation}
and we use the extended threshold
\begin{equation}\label{eq:axis-B}
 B(\tau)=\sup\{s>0:P(\tau,s,0)=K-s\},\qquad b(\tau,0)=B(\tau),\qquad\tau>0.
\end{equation}
The set in \eqref{eq:axis-B} is closed by continuity of $P$, so the
supremum is attained and value matching holds at $B(\tau)$. The
probabilistic properties of Proposition~\ref{probprops} hold also at
$y=0$. In particular $0<B(\tau)<K$, where positivity follows from
$B(\tau)\ge b(\tau,y)$ for every $y>0$ and the strict inequality at the
strike follows from the strictly positive European put started at zero
variance. The function $R$ is nonnegative, nondecreasing in $\tau$, $s$
and $y$, convex and $1$-Lipschitz in $s$, and $R(\tau,0,y)=0$. In this section $\nu$ denotes the CIR index,
spatial normals are written $n$ as in Section~\ref{sec:geometry}, and
the letters $M$, $M_1$ and $M_*$ denote positive constants, not the
matrix of \eqref{matrixM}.

\subsection{Price estimates at zero variance}

The next estimate uses the local time of the stock at the strike.  It does
not use elliptic or parabolic estimates at the degenerate boundary.

\begin{lemma}\label{lem:axis-time-lipschitz}
Fix $0<\tau_-<\tau_+<\infty$.  There is a constant $C$, depending on this
interval and the model parameters, such that, for all $s>0$, $y\ge0$, and
$\tau_-\le t\le t+h\le\tau_+$,
\begin{equation}\label{eq:axis-time-lipschitz}
 0\le P(t+h,s,y)-P(t,s,y)\le C(1+y)h.
\end{equation}
In particular, the price is locally Lipschitz in remaining maturity,
uniformly in stock price and on bounded variance intervals, including zero.
\end{lemma}

\begin{proof}
Write $I_u=\int_0^uY_v\,dv$ and
$W^S=\rho W+\sqrt{1-\rho^2}\,W^\perp$, where $W^\perp$ is independent
of $W$.  Conditional on the path of $W$, the variable $\log S_u$ is
Gaussian with variance $(1-\rho^2)I_u$ and mean
\[
 \log s+(r-q)u-\tfrac12 I_u
       +\frac{\rho}{\sigma}(Y_u-y-\kappa\theta u+\kappa I_u).
\]
Here $I_u>0$ almost surely for $u>0$, also when $y=0$.  Its conditional
stock density therefore satisfies
\begin{equation}\label{eq:conditional-stock-density}
 p_u(z\mid W)\le
 \frac{1}{z\sqrt{2\pi(1-\rho^2)I_u}},\qquad z>0.
\end{equation}

We first bound the inverse moment of $I_u$.  Couple the CIR solutions
with their common Brownian motion.  Pathwise uniqueness holds for the
CIR equation by the Yamada--Watanabe theorem \cite{YW}, and the
comparison theorem for one-dimensional diffusions \cite[Chapter VI,
Theorem 1.1]{IW} gives $Y^y\ge Y^0$, so
$I_u^y\ge I_{\tau_-}^0$ whenever $u\ge\tau_-$.  Put
$d_\lambda=\sqrt{\kappa^2+2\sigma^2\lambda}$.  The CIR affine transform,
obtained by substituting an exponential-affine function into the backward
equation, gives, for every $T>0$,
\begin{equation}\label{eq:integrated-cir-laplace}
 \E_0 e^{-\lambda I_T}
 =\left[
 \frac{2d_\lambda e^{(\kappa+d_\lambda)T/2}}
 {(d_\lambda+\kappa)(e^{d_\lambda T}-1)+2d_\lambda}
 \right]^\nu.
\end{equation}
It follows from the exponential-affine ansatz $F(T,y)=A(T)e^{-D(T)y}$ with
$D'=\lambda-\kappa D-\sigma^2D^2/2$, $D(0)=0$, $A'/A=-\kappa\theta D$,
$A(0)=1$, see also \cite{PitmanYor,Alfonsi}.
Its right-hand side is at most $C_T e^{-c_T\sqrt\lambda}$ for large
$\lambda$, with $c_T>0$.  Tonelli's theorem consequently yields
\begin{equation}\label{eq:integrated-cir-inverse}
 \E_0[I_T^{-1}]
 =\int_0^\infty\E_0e^{-\lambda I_T}\,d\lambda<\infty.
\end{equation}
The CIR moment equations give
$\sup_{0\le u\le\tau_+}\E_yY_u^2\le C(1+y^2)$.
Comparison and Cauchy--Schwarz now imply
\begin{equation}\label{eq:local-time-rate-bound}
 \sup_{\tau_-\le u\le\tau_+}
 \E_y\left[\frac{Y_u}{\sqrt{I_u}}\right]\le C(1+y).
\end{equation}

Let $L^K(S)$ denote the symmetric semimartingale local time at $K$,
see \cite[Chapter VI]{RY}.
The occupation formula, \eqref{eq:conditional-stock-density}, and
\eqref{eq:local-time-rate-bound} give
\begin{equation}\label{eq:strike-local-time-increment}
 \E_{s,y}[L^K_{t+h}(S)-L^K_t(S)]\le C(1+y)h.
\end{equation}
To see this without a joint transition density of $(S,Y)$, bound the
occupation approximation
$(2\varepsilon)^{-1}\int_t^{t+h}\mathbf1_{\{|S_u-K|<\varepsilon\}}S_u^2Y_u\,du$,
$0<\varepsilon<K/2$, by conditioning on $W$: since
$\int_{K-\varepsilon}^{K+\varepsilon}z\,dz=2K\varepsilon$, its expectation is at
most $K(2\pi(1-\rho^2))^{-1/2}\int_t^{t+h}\E_y[Y_uI_u^{-1/2}]\,du$, and the
occupation limit together with Fatou's lemma proves
\eqref{eq:strike-local-time-increment}. The conditional density also shows
that $S$ spends zero Lebesgue time at $K$, so the left, right and symmetric
local times at this level coincide.

Set $Z_u=e^{-ru}(K-S_u)^+$.  It\^o--Tanaka's formula reads
\[
 Z_u=Z_0+\mathcal M_u-
 \int_0^u e^{-rv}(rK-qS_v)\mathbf1_{\{S_v<K\}}\,dv
 +\frac12\int_0^u e^{-rv}\,dL_v^K(S).
\]
The process $\mathcal M$ is a true square-integrable martingale on each bounded
time interval: its integrand has absolute value at most $K\sqrt{Y_v}$.
The finite-variation drift displayed above is nonpositive because
$0\le q\le r$.  For every stopping time $\eta\le t+h$, optional
sampling at $\eta$ and $\eta\wedge t$ gives
\[
 \E Z_\eta-\E Z_{\eta\wedge t}
 \le\tfrac12\E[L^K_{t+h}(S)-L^K_t(S)].
\]
Since $\E Z_{\eta\wedge t}\le P(t,s,y)$, taking the supremum over
$\eta$ proves the upper bound in \eqref{eq:axis-time-lipschitz}.
The lower bound is time monotonicity.
\end{proof}

The definition \eqref{eq:axis-B} of $B$ makes at this stage no
assertion about the limit of $b(t,y)$ as $y\searrow0$. That limit is
identified in Theorem~\ref{thm:axis-cross-maturity} below.

\begin{proposition}\label{prop:axis-linear-premium}
Fix $0<\tau_-<\tau_+<\infty$.  There are $C,y_0>0$ such that
\begin{equation}\label{eq:axis-linear-premium}
 0\le R(t,s,y)\le Cy,
 \qquad \tau_-\le t\le\tau_+,\quad 0<s\le B(t),\quad 0\le y\le y_0.
\end{equation}
Consequently,
\begin{equation}\label{eq:axis-global-premium}
 0\le R(t,s,y)\le Cy+(s-B(t))^+
\end{equation}
on the same time and variance intervals and for every $s>0$.
\end{proposition}

\begin{proof}
Start the process at $(B(t),0)$.
For $0<h\le\tau_-/2$, contact and the dynamic programming inequality
give
\[
 \E[e^{-rh}P(t-h,S_h,Y_h)]\le K-B(t).
\]
The discounted stock is a nonnegative supermartingale, including
the dividend adjustment. Hence
\begin{align*}
 \E[e^{-rh}R(t-h,S_h,Y_h)]
 &\le K(1-e^{-rh})-B(t)(1-e^{-qh})\\
 &\le rKh.
\end{align*}
Apply Lemma~\ref{lem:axis-time-lipschitz} on
$[\tau_-/2,\tau_+]$.  Its factor $1+y$ is useful because $Y_h$ is
not bounded.  It gives
\[
 0\le R(t,S_h,Y_h)-R(t-h,S_h,Y_h)\le C(1+Y_h)h.
\]
As $\E_0Y_h=\theta(1-e^{-\kappa h})$, it follows that
\begin{equation}\label{eq:axis-dpp-premium}
 \E[e^{-rh}R(t,S_h,Y_h)]\le C h,
\end{equation}
uniformly for $\tau_-\le t\le\tau_+$.

The CIR moment equations at zero give
\[
 \bar y_h:=\E_0Y_h=\theta(1-e^{-\kappa h}),\qquad
 \E_0Y_h^2=\theta\left(\theta+\frac{\sigma^2}{2\kappa}\right)
                        (1-e^{-\kappa h})^2.
\]
The Paley--Zygmund inequality therefore gives
\[
 \PP_0(Y_h\ge \bar y_h/2)\ge\frac{\nu}{4(\nu+1)}=:p_0>0.
\]
Choose $h=4y/a$ and reduce $y_0$ so that
$h\le\min\{1,1/\kappa,\tau_-/2\}$.
Then $\bar y_h\ge ah/2=2y$, and $\PP_0(Y_h\ge y)\ge p_0$.
Moreover, $\xi_h=\log(S_h/B(t))$ satisfies
\[
 \E|\xi_h|
 \le |r-q|h+\frac a4h^2+\sqrt{a/2}\,h\le C_1h.
\]
This follows from $\E_0Y_u\le au$ and It\^o's isometry, and the constant
is independent of $t$ and $B(t)$.  Choose $M_1\ge2C_1/p_0$.
The union bound then yields
\[
 \PP(S_h\ge B(t)e^{-M_1h},\ Y_h\ge y)\ge p_0/2.
\]
No independence between these two events is used.  Monotonicity in
$s,y$ and \eqref{eq:axis-dpp-premium} give
\[
 R(t,B(t)e^{-M_1h},y)\le \frac{2e^{rh}}{p_0}C h.
\]
The stock Lipschitz bound and $B(t)<K$ imply
\[
 R(t,B(t),y)
 \le \frac{2e^{rh}}{p_0}C h+K(1-e^{-M_1h})\le C' y.
\]
Monotonicity proves \eqref{eq:axis-linear-premium}, and the same
Lipschitz bound proves \eqref{eq:axis-global-premium}.
\end{proof}

\subsection{Lipschitz continuity in remaining maturity, uniformly in the variance}
\label{subsec:axis-lipschitz}

The estimates of this subsection are stated for an arbitrary initial
variance $y\ge0$. At $y=0$ they concern the axis threshold $B$, and for
$y>0$ they give a bound on $b_\tau$ that does not deteriorate as the
variance tends to zero. Recall that for $y\ge0$ the stock started at
$(s,y)$ has the form $S^{s,y}_t=sN^y_t$, where $N^y$ does not depend
on $s$.

\begin{lemma}\label{lem:axis-strict-contact-margin}
Fix $\tau>0$ and $y\ge0$, and write $b=b(\tau,y)$. For $0<s<b$ and
every stopping time $\eta\le\tau$, with initial state $(s,y)$,
\[
 \mathbb E[e^{-r\eta}g(S^{s,y}_\eta)]-g(s)
 \le-(b-s)\,\frac{1-e^{-r\tau}}\tau\,\mathbb E\eta .
\]
\end{lemma}
\begin{proof}
Convexity of $g$ and $g(0)=K$ give
\[
 g(sN^y_\eta)\le\frac{s}{b}g(bN^y_\eta)
              +\left(1-\frac{s}{b}\right)K.
\]
The point $(\tau,b,y)$ belongs to the contact set, so
$\mathbb E[e^{-r\eta}g(bN^y_\eta)]\le P(\tau,b,y)=g(b)$. Moreover
$g(s)=(s/b)g(b)+(1-s/b)K$. Subtraction therefore gives
\[
 \mathbb E[e^{-r\eta}g(sN^y_\eta)]-g(s)
 \le-K\Bigl(1-\frac sb\Bigr)\mathbb E(1-e^{-r\eta}).
\]
Since $b\le K$, the prefactor is at least $b-s$, and the concavity of
$z\mapsto1-e^{-rz}$ on $[0,\tau]$ yields the result.
\end{proof}

\begin{lemma}\label{lem:weighted-horizon-extension}
For every compact interval $I\Subset(0,\infty)$ there are
$C_I,h_I>0$ with the following property. Start the process at
$(s,y)$ with $s>0$ and $y\ge0$, let $\tau\in I$, $0<h<h_I$, and let
$\eta\le\tau+h$ be a stopping time. Put $\eta_1=\eta\wedge\tau$. Then
\[
 \mathbb E[e^{-r\eta}g(S_\eta)]
 -\mathbb E[e^{-r\eta_1}g(S_{\eta_1})]
 \le C_I(1+y)h\,\mathbb E\eta_1 .
\]
The constant is independent of the initial stock and the stopping time.
\end{lemma}
\begin{proof}
Let $L^K_t$ denote the semimartingale local time of the stock at $K$,
with the convention that its coefficient in Tanaka's formula is
$1/2$. The discounted payoff has differential
\[
 d(e^{-rt}g(S_t))
 =-e^{-rt}(rK-qS_t)\mathbf1_{\{S_t<K\}}\,dt
  -e^{-rt}S_t\sqrt{Y_t}\mathbf1_{\{S_t<K\}}\,dW^S_t
  +\tfrac12e^{-rt}\,dL^K_t.
\]
The drift is nonpositive because $0\le q\le r$, and the stochastic
integral is a true martingale on bounded time intervals, since its
expected quadratic variation is at most
$K^2\mathbb E\int_0^{\tau+h}Y_t\,dt$.
On $\{\eta\le\tau\}$ the rewards at $\eta$ and $\eta_1$ coincide.
On the complementary event Tanaka's formula and positivity of local
time give, after taking expectations,
\begin{align*}
 &\mathbb E[e^{-r\eta}g(S_\eta)]
  -\mathbb E[e^{-r\eta_1}g(S_{\eta_1})]\\
 &\qquad\le\tfrac12\mathbb E[
       \mathbf1_{\{\eta>\tau\}}(L^K_{\tau+h}-L^K_\tau)]\\
 &\qquad\le\tfrac12\mathbb E[
       \mathbf1_{\{\eta>\tau/2\}}(L^K_{\tau+h}-L^K_\tau)].
\end{align*}
The event in the last line belongs to $\mathcal F_{\tau/2}$.
Estimate \eqref{eq:strike-local-time-increment} states that, for any
fixed $0<t_0<t_1<\infty$,
\[
 \mathbb E_{s',y'}[L^K_{u+h}-L^K_u]\le C(1+y')h,
 \qquad t_0\le u\le t_1,
\]
uniformly over $s'>0$, $y'\ge0$ and sufficiently small $h$.
Apply the strong Markov property at $\tau/2$ and use this estimate
with the positive delay $u=\tau/2$. The preceding reward difference
is at most
\[
 Ch\,\mathbb E\bigl[
       \mathbf1_{\{\eta>\tau/2\}}(1+Y_{\tau/2})\bigr].
\]
Set $T=\tau/2$ and $\eta_2=\eta\wedge T$, so that $\eta_2\le\eta_1$. We have
\[
 \mathbb P(\eta>T)\le\frac{1}{T}\mathbb E\eta_2
                       \le\frac{2}{\tau}\mathbb E\eta_1.
\]
For the variance term we use a Lyapunov function. Put
\[
 \psi(t,z)=\frac{2(1+y)t}{T}+\frac{2(z-y)^2}{1+y},
 \qquad t\ge0,\ z\ge0 .
\]
Then $\psi\ge0$, $\psi(0,y)=0$, and $\psi(T,z)\ge z$ for every
$z\ge0$. Indeed, if $z\le2(1+y)$ the first term already dominates $z$,
and if $z>2(1+y)$ then $z-y>z/2$ and $2(z-y)^2>z^2/2\ge z(1+y)$.
With the CIR generator $\mathcal G=\varsigma z\partial_{zz}+(a-\kappa
z)\partial_z$,
\begin{align*}
 (\partial_t+\mathcal G)\psi
 &=\frac{2(1+y)}{T}+\frac{2\sigma^2z+4(a-\kappa z)(z-y)}{1+y}\\
 &\le\frac{2(1+y)}{T}+\frac{(4a+4\kappa y+2\sigma^2)^2}{16\kappa(1+y)}
 \le C(1+y)\Bigl(1+\frac1\tau\Bigr),
\end{align*}
because the numerator of the middle term is a concave quadratic in
$z$. It\^o's formula at the bounded time $\eta_2$, where the martingale
term has zero expectation since all CIR moments are finite on bounded
intervals, gives
\[
 \mathbb E[Y_T\mathbf1_{\{\eta>T\}}]
 \le\mathbb E\,\psi(\eta_2,Y_{\eta_2})
 \le C(1+y)\Bigl(1+\frac1\tau\Bigr)\mathbb E\eta_2
 \le C(1+y)\Bigl(1+\frac1\tau\Bigr)\mathbb E\eta_1 ,
\]
where the first inequality uses $\eta_2=T$ on $\{\eta>T\}$, the bound
$\psi(T,\cdot)\ge z$ there, and $\psi\ge0$ elsewhere. Combining the
displays proves the claim uniformly on compact positive maturity
intervals.
\end{proof}

\begin{proposition}\label{prop:axis-maturity-lipschitz}
For each compact interval $I\Subset(0,\infty)$ there are $C_I,h_I>0$
such that
\begin{equation}\label{eq:axis-uniform-lipschitz}
 0\le b(\tau,y)-b(\tau+h,y)\le C_I(1+y)h
\end{equation}
whenever $\tau\in I$, $y\ge0$ and $0<h<h_I$, where $b(\tau,0)=B(\tau)$.
In particular the zero-variance threshold $B$ is locally Lipschitz
continuous on $(0,\infty)$, and for every $Y_*>0$,
\begin{equation}\label{eq:axis-btau-bound}
 -C_I(1+Y_*)\le b_\tau(\tau,y)\le0,\qquad \tau\in I,\ 0<y\le Y_*.
\end{equation}
\end{proposition}
\begin{proof}
Fix $\tau\in I$, $y\ge0$, $0<h<h_I$, and let $s<b(\tau,y)$. For a
stopping time $\eta\le\tau+h$ put $\eta_1=\eta\wedge\tau$. Combining
Lemmas~\ref{lem:axis-strict-contact-margin}
and~\ref{lem:weighted-horizon-extension} for the process started at
$(s,y)$ gives
\[
 \mathbb E[e^{-r\eta}g(S^{s,y}_\eta)]-g(s)
 \le\bigl[-c_I(b(\tau,y)-s)+C_I(1+y)h\bigr]\mathbb E\eta_1,
\]
where $c_I>0$ is the positive minimum of $(1-e^{-r\tau})/\tau$ on $I$
and is independent of $s,y,\tau,h$.
Choose $D_I>C_I/c_I$. If $s=b(\tau,y)-D_I(1+y)h>0$, every admissible
stopping time for the maturity $\tau+h$ has expected reward at most
$g(s)$, and immediate exercise attains it. Hence $(\tau+h,s,y)$ is a
contact point and $b(\tau+h,y)\ge b(\tau,y)-D_I(1+y)h$.
If $b(\tau,y)-D_I(1+y)h\le0$, this inequality is automatic.
The opposite inequality follows from maturity monotonicity. This
proves \eqref{eq:axis-uniform-lipschitz} with $C_I$ replaced by $D_I$.
Applying the estimate to the smaller of any two nearby positive
maturities proves local Lipschitz continuity of $B$, and for $y>0$ the
difference quotients of the smooth function $b(\cdot,y)$ give
\eqref{eq:axis-btau-bound}.
\end{proof}

\begin{remark}\label{rem:axis-btau}
The bound \eqref{eq:axis-btau-bound} is the only estimate of this paper on the
maturity derivative of the boundary that survives at the axis. The boundary
surface moves with bounded speed in remaining maturity down to zero variance,
even though its variance slope $b_y$ blows up there by
Proposition~\ref{prop:finite-axis-normals}. By \eqref{eq:axis-uniform-lipschitz}
the functions $b_\tau(\cdot,y)$ are bounded in $L^\infty_{\loc}$ uniformly in
small $y$ and converge to $B'$ in the weak-$*$ sense as $y\searrow0$, while
pointwise convergence is not claimed.
\end{remark}

\subsection{Matching with the axis and a quantitative boundary modulus}

For $0<v\le1$, define
\[
 \omega_\nu(v)=
 \begin{cases}
 v^{\nu/(\nu+1)},&0<\nu<1,\\
 \{v\log(e/v)\}^{1/2},&\nu=1,\\
 v^{1/2},&\nu>1.
 \end{cases}
\]
The matching estimate below allows a positive change in remaining
maturity.  Keeping this change explicit avoids assuming time regularity
of the endpoint $B$.

\begin{lemma}\label{lem:axis-cir-exit}
Write $\mathcal G=\varsigma y\partial_{yy}+(a-\kappa y)\partial_y$
for the CIR generator.
Fix $L_0\le\min\{1,a/(2\kappa)\}$.  For
$0<v\le L/8$ and $L\le L_0$, let $m=m_{v,L}$ solve
\[
 \mathcal G m=-1\quad\hbox{in }(v,L),\qquad m(v)=m(L)=0.
\]
There are constants $c,C>0$, independent of $v,L$, such that
\[
 0\le m\le CL,\qquad y|m'(y)|\le CL,
 \qquad m(2v)\ge cL\Phi_\nu(v/L),
\]
where
\[
 \Phi_\nu(z)=
 \begin{cases}
 z^{1-\nu},&0<\nu<1,\\
 1/\log(1/z),&\nu=1,\\
 1,&\nu>1.
 \end{cases}
\]
\end{lemma}

\begin{proof}
The maximum principle and the supersolution $2(L-y)/a$ give
$0\le m\le2L/a$.  Set $\lambda=\kappa/\varsigma$,
$p(y)=y^{-\nu}e^{\lambda y}$, and
$J(y)=\varsigma^{-1}\int_0^y z^{\nu-1}e^{-\lambda z}\,dz$.
An integrating factor gives
\[
 m'(y)=p(y)(C_{v,L}-J(y)),\qquad
 C_{v,L}=\frac{\int_v^L p(z)J(z)\,dz}{\int_v^L p(z)\,dz}.
\]
If $\nu\le1$, the bounds $J(y)\le y^\nu/a$ and
$C_{v,L}\le J(L)$ show $y|m'(y)|\le CL$.
If $\nu>1$, the numerator is at most $CL$ and the denominator
is at least $c_\nu v^{1-\nu}$, and hence
$C_{v,L}\le CLv^{\nu-1}$, which gives the same derivative bound.

On $(L/4,3L/4)$, the test function
$(y-L/4)(3L/4-y)/L$ has uniformly bounded image under
$\mathcal G$.  Comparison with a fixed multiple of this function
gives $m(L/2)\ge cL$.  The strong Markov property and the CIR scale
density $p$ then yield
\[
 m(2v)\ge m(L/2)\PP_{2v}(T_{L/2}<T_v)
 =m(L/2)\frac{\int_v^{2v}p(z)\,dz}
                       {\int_v^{L/2}p(z)\,dz}.
\]
The factor $e^{\lambda z}$ is bounded above and below on $(0,L_0)$.
Integrating $z^{-\nu}$ in the three cases proves the claim.
\end{proof}

\begin{theorem}\label{thm:axis-cross-maturity}
For each $t>0$ and $h>0$, there are constants $C_{t,h},v_{t,h}>0$
such that
\begin{equation}\label{eq:axis-cross-maturity}
 \left[\log\frac{B(t+2h)}{b(t,v)}\right]^+
 \le C_{t,h}\omega_\nu(v),\qquad 0<v<v_{t,h}.
\end{equation}
The interior trace
\[
 b_*(t):=\lim_{v\searrow0}b(t,v)
\]
exists, and it satisfies the exact identity
\begin{equation}\label{eq:axis-right-trace}
 b_*(t)=B(t+):=\lim_{h\searrow0}B(t+h).
\end{equation}
The function $B$ is nonincreasing and left continuous on $(0,\infty)$.
Proposition~\ref{prop:axis-maturity-lipschitz} gives right continuity as well.
Consequently $b(t,v)\to B(t)$ as $v\searrow0$ for every $t>0$,
and the extension $b(t,0)=B(t)$ is jointly continuous at every
$(t,0)$ with $t>0$.
\end{theorem}

\begin{proof}
Use the logarithmic gap $w$ from \eqref{logdefs} and put
$H=w_x-w$.  Stock convexity and $R(t,0,y)=0$ give
$H=e^xR_s-R\ge0$.  In the continuation region,
\begin{equation}\label{eq:axis-H-parabolic}
 (\partial_t-\Lh+r)H=rK.
\end{equation}
Indeed $(\partial_t-\Lh+r)w=-rK+qe^x$, and the coefficients
of $\Lh$ are independent of $x$.  The time derivative in
\eqref{eq:axis-H-parabolic} is retained throughout the argument.

Fix $t,h$, write $x_0=\log B(t+2h)$, and choose
$0<\delta_*\le\min\{1,4L_0,h\}$.  Reduce it if necessary so that
the premium estimate of Proposition~\ref{prop:axis-linear-premium}
applies for $0\le y\le\delta_*/4$ and remaining times in $[t,t+2h]$.
For a given $v>0$, set
\[
 d=\left[x_0-\log b(t,v)\right]^+,
 \qquad \delta=\min\{d,\delta_*\},\qquad L=\delta/4,
 \qquad x_c=x_0-\delta/2.
\]
There is nothing to prove when $\delta=0$.  If $L<8v$, then
$\delta<32v$ already gives the required estimate for small variances.
Assume henceforth $8v\le L$.

The cylinder
\[
 [t,t+h]\times\overline Q,\qquad
 Q=(x_c-L,x_c+L)\times(v,L),
\]
lies strictly in the continuation region.  In fact, time and variance
monotonicity give $b(u,y)\le b(t,v)$ for $u\ge t,y\ge v$, whereas
$x\ge x_0-3\delta/4>x_0-d=\log b(t,v)$ throughout its spatial
closure.

Let $m=m_{v,L}$ be as in Lemma~\ref{lem:axis-cir-exit}, and set
$\phi(x)=1-(x-x_c)^2/L^2$.  A direct computation gives
\[
 (\Lh-r)(\phi m)
 =-\phi+\frac y2\phi''m+(r-q-y/2)\phi'm
       +\rho\sigma y\phi'm'-r\phi m.
\]
The bounds of Lemma~\ref{lem:axis-cir-exit} and $y\le L$ show that the absolute value of
this expression is bounded by a constant independent of $v,L,x_c$.
Thus, after division by a fixed constant, the function
$\Lambda(x,y)=c_0\phi(x)m(y)$ satisfies
\[
 \Lambda\ge0,\quad \Lambda|_{\partial Q}=0,\quad
 (\Lh-r)\Lambda\ge-1,\quad
 \Lambda\le CL,\quad \Lambda(x_c,2v)\ge cL\Phi_\nu(v/L).
\]
For $t\le u\le t+h$, define
$\varpi(u)=(u-t)/h$ and $\Upsilon(u,x,y)=\varpi(u)\Lambda(x,y)$.
Since $L\le h/4$,
\[
 (\partial_u-\Lh+r)\Upsilon
 =\Lambda/h-\varpi(\Lh-r)\Lambda\le C_0.
\]
The function $H-(rK/C_0)\Upsilon$ is nonnegative on the parabolic
boundary of this cylinder and has nonnegative image under
$\partial_u-\Lh+r$.  The comparison principle therefore gives
\begin{equation}\label{eq:axis-H-lower}
 H(t+h,x_c,2v)\ge cL\Phi_\nu(v/L).
\end{equation}

For $t\le u\le t+h$, one has $B(t+2h)\le B(u)$.
Proposition~\ref{prop:axis-linear-premium} consequently gives
$R(u,B(t+2h),y)\le Cy$.  The secant inequality for the convex
function $s\mapsto R(u,s,y)$ implies, when $x\le x_0-\delta/4$,
\[
 0\le H(u,x,y)\le e^xR_s(u,e^x,y)
 \le\frac{e^x}{B(t+2h)-e^x}R(u,B(t+2h),y)
 \le C\frac y\delta.
\]
Combining this estimate at $(t+h,x_c,2v)$ with
\eqref{eq:axis-H-lower} gives
\[
 C\frac v\delta\ge cL\Phi_\nu(v/L).
\]
As $L=\delta/4$, this yields, respectively,
\[
 \delta^{\nu+1}\le Cv^\nu,\qquad
 \delta^2\le Cv\log(e/v),\qquad
 \delta^2\le Cv
\]
in the cases $\nu<1$, $\nu=1$, and $\nu>1$.
The alternative $\delta<32v$ obeys the same estimates.
The resulting upper bound tends to zero, so for sufficiently small
$v$ it is smaller than $\delta_*$ and hence $\delta=d$.
This proves \eqref{eq:axis-cross-maturity}.

Variance monotonicity guarantees the existence of $b_*(t)$.
For $s<b_*(t)$, sufficiently small positive variances are contact
states, and continuity of $P$ at $y=0$ gives $s\le B(t)$.
Thus $b_*(t)\le B(t)$.  Letting $v\searrow0$ in
\eqref{eq:axis-cross-maturity} and then $h\searrow0$ gives
\begin{equation}\label{eq:axis-trace-sandwich}
 B(t+)\le b_*(t)\le B(t).
\end{equation}
The interior smoothness theorem shows that every function
$t\mapsto b(t,v)$, $v>0$, is continuous.  Therefore
$b_*(t)=\sup_{v>0}b(t,v)$ is lower semicontinuous.  Since it is
nonincreasing, it is right continuous.  Apply the upper inequality
in \eqref{eq:axis-trace-sandwich} at $t+h$ and let $h\searrow0$:
\[
 b_*(t)=\lim_{h\searrow0}b_*(t+h)
       \le\lim_{h\searrow0}B(t+h)=B(t+).
\]
Together with the other inequality, this proves
\eqref{eq:axis-right-trace}.

The function $B$ is nonincreasing by time monotonicity of the price.
If $t_n\nearrow t$, then $B(t_n)$ decreases to a limit $\ell\ge B(t)$.
Continuity of $P$ and value matching give $P(t,\ell,0)=K-\ell$,
so $\ell\le B(t)$ by the definition of $B$.  Hence $B$ is left
continuous.  Proposition~\ref{prop:axis-maturity-lipschitz} gives right
continuity as well.
Thus $B(t+)=B(t)$, and \eqref{eq:axis-right-trace} proves matching
at every time.

Finally, fix any $t_0>0$.  For $\varepsilon>0$
small enough and $|t-t_0|<\varepsilon$, monotonicity gives
\[
 b(t_0+\varepsilon,v)\le b(t,v)
 \le b(t_0-\varepsilon,v),\qquad v>0.
\]
Take $v\searrow0$, use \eqref{eq:axis-right-trace}, and then let
$\varepsilon\searrow0$.  Both outer limits tend to $B(t_0)$.
The same conclusion holds along points on the axis by continuity of
$B$ at $t_0$.  This proves joint continuity there.
\end{proof}

\begin{corollary}\label{cor:axis-uniform-matching}
For every compact interval $I\Subset(0,\infty)$,
$\sup_{\tau\in I}\abs{b(\tau,y)-B(\tau)}\to0$ as $y\searrow0$.
\end{corollary}
\begin{proof}
For any sequence $y_n\searrow0$, the continuous functions
$b(\cdot,y_n)$ increase to the continuous function $B$ on $I$ by
Theorem~\ref{thm:axis-cross-maturity} and
Proposition~\ref{prop:axis-maturity-lipschitz}, so Dini's theorem gives
uniform convergence along the sequence. Monotonicity in $y$ gives the
same conclusion without restricting to a sequence.
\end{proof}

The local Lipschitz estimate for $B$ converts the preceding comparison
into an estimate at a fixed remaining maturity, with the same function
$\omega_\nu$ as in the cross-maturity estimate.

\begin{theorem}\label{thm:axis-quantitative-modulus}
Let $I\Subset(0,\infty)$ be a compact interval of remaining maturities.
There exist $C_I,v_I>0$ such that
\begin{equation}\label{eq:axis-fixed-maturity-modulus}
 0\le\log\frac{B(t)}{b(t,v)}
 \le C_I\omega_\nu(v),
 \qquad t\in I,\quad 0<v<v_I.
\end{equation}
In particular,
\begin{equation}\label{eq:axis-stock-fixed-maturity-modulus}
 0\le B(t)-b(t,v)\le K C_I\omega_\nu(v)
\end{equation}
uniformly for $t\in I$.  These are upper estimates and do not
assert sharp asymptotic equivalents.
\end{theorem}

\begin{proof}
Enlarge $I$ to a fixed compact interval of positive maturities so
that the small positive shifts used below remain in that interval.
The endpoint is bounded away from zero there, because
$B(u)\ge b(u,v_0)>0$ for any fixed $v_0>0$ and the interior
threshold is continuous.  The local Lipschitz estimate of
Proposition~\ref{prop:axis-maturity-lipschitz} therefore implies, for
$\beta_0(u)=\log B(u)$,
\begin{equation}\label{eq:axis-log-time-holder}
 0\le\beta_0(t)-\beta_0(t+2h)\le Ah
\end{equation}
for $t\in I$ and all sufficiently small $h>0$, with $A$ uniform.

Let $L_0$ be as in Lemma~\ref{lem:axis-cir-exit} and let $y_0$
be the uniform variance cutoff in
Proposition~\ref{prop:axis-linear-premium} on the enlarged time
interval.  Choose fixed $0<\delta_0\le1$ and $c_0>0$ so small that
$Ac_0\le1/4$, $c_0\delta_0\le\min\{L_0,y_0,1\}$, and all
time shifts $2c_0\delta_0$ are permitted in
\eqref{eq:axis-log-time-holder}.
Fix $t\in I$ and $v>0$, and set
\begin{gather*}
 d=\log\frac{B(t)}{b(t,v)},\quad
 \delta=\min\{d,\delta_0\},\quad
 h=L=c_0\delta,\quad \ell=\delta/8,\\
 x_0=\beta_0(t),\quad x_c=x_0-\delta/2.
\end{gather*}
When $\delta=0$ the conclusion is immediate.  If $L<8v$, then
$\delta<8v/c_0$, which is stronger than the desired estimate.
We may therefore assume $8v\le L$.

Consider the spatial rectangle
\[
 Q=(x_c-\ell,x_c+\ell)\times(v,L)
\]
and the time interval $[t,t+h]$.  The entire closed cylinder lies
in continuation: for $u\ge t$ and $y\ge v$,
$b(u,y)\le b(t,v)$, whereas its smallest logarithmic stock coordinate is
$x_0-5\delta/8>x_0-d=\log b(t,v)$.
Furthermore, \eqref{eq:axis-log-time-holder} gives
\[
 \beta_0(t+2h)\ge x_0-\delta/4.
\]
Every stock coordinate in the rectangle is thus below the future
axis threshold, with a logarithmic margin at least $\delta/8$.

As in the proof of Theorem~\ref{thm:axis-cross-maturity}, let
$\phi(x)=1-(x-x_c)^2/\ell^2$ and take a fixed multiple
$\Lambda(x,y)$ of $\phi(x)m_{v,L}(y)$ such that
\[
 \Lambda\ge0,\quad \Lambda|_{\partial Q}=0,\quad
 (\Lh-r)\Lambda\ge-1,\quad \Lambda\le C L,
 \quad \Lambda(x_c,2v)\ge cL\Phi_\nu(v/L).
\]
The derivative bounds for $m$ and the relations $L=c_0\delta$ and
$\ell=\delta/8$ show that all constants are independent of $t,v,\delta$.
Indeed, the stock diffusion, stock drift, and mixed terms in
$(\Lh-r)(\phi m)$ are bounded by $CL^2/\delta^2$, $CL/\delta$,
and $CL/\delta$, respectively.
Set $\Upsilon(u,x,y)=((u-t)/h)\Lambda(x,y)$.  As $L/h=1$,
\[
 (\partial_u-\Lh+r)\Upsilon\le1+CL/h\le C_1.
\]
For $H=w_x-w\ge0$, equation
\eqref{eq:axis-H-parabolic} and comparison on the cylinder imply
\[
 H(t+h,x_c,2v)
 \ge\frac{rK}{C_1}\Lambda(x_c,2v)
 \ge c\delta\Phi_\nu(v/L).
\]

On the other hand, $B(t+2h)\le B(u)$ for $u\le t+h$.
The linear premium estimate and stock convexity therefore give,
on this rectangle,
\begin{align*}
 H(u,x,y)
 &\le\frac{e^x}{B(t+2h)-e^x}
                   R(u,B(t+2h),y)\\
 &\le C y/\delta.
\end{align*}
Evaluate this at $(t+h,x_c,2v)$ and combine it with the lower
estimate.  Since $L=c_0\delta$, the resulting inequality is
\begin{equation}\label{eq:axis-fixed-gap-estimate}
 \delta^2\Phi_\nu\bigl(v/(c_0\delta)\bigr)\le C v.
\end{equation}
If $0<\nu<1$, it gives $\delta^{\nu+1}\le C v^\nu$.
If $\nu=1$, it gives
$\delta^2\le Cv\log(L/v)\le Cv\log(e/v)$.
If $\nu>1$, it gives $\delta^2\le Cv$.
The alternative $\delta<8v/c_0$ satisfies the same bounds after
enlarging $C$.  Thus
$\delta\le C\omega_\nu(v)$ uniformly on $I$.
For sufficiently small $v$, the right-hand side is less than
$\delta_0$, so the truncation is inactive and $\delta=d$.
This proves \eqref{eq:axis-fixed-maturity-modulus}.
Finally,
$B(t)-b(t,v)=B(t)(1-e^{-d})\le Kd$, proving
\eqref{eq:axis-stock-fixed-maturity-modulus}.
\end{proof}

\subsection{Linear nondegeneracy and separation at zero variance}

Proposition~\ref{prop:axis-linear-premium} and
Lemma~\ref{lem:axis-time-lipschitz} give, for $\tau$ in a compact
interval of positive maturities and $0\le y\le y_0$,
\begin{equation}\label{eq:axis-w-bounds}
 0\le w(\tau,x,y)\le Cy+(e^x-B(\tau))^+,
 \qquad 0\le P_\tau\le C(1+y),
\end{equation}
where the second inequality holds almost everywhere and in the
corresponding Lipschitz sense, and where $w$ is the logarithmic gap
of \eqref{logdefs}.

The lifting of Lemma~\ref{liftlemma} remains uniform as the discount tends to
zero, which is needed below because the rescaled discount is $\varepsilon r$.
Let $A_\varepsilon$ be uniformly elliptic on a fixed positive-variance compact
set, let $B_\varepsilon$ be uniformly bounded together with its derivatives, and
let $c_\varepsilon=\varepsilon r\in[0,r]$. Choose $M_*$ large and independent
of $\varepsilon$, and let $\phi_\varepsilon$ solve
$\phi_\varepsilon''=-c_\varepsilon\phi_\varepsilon$ with $\phi_\varepsilon(0)=1$
and $\phi_\varepsilon'(0)=-M_*$. On a sufficiently short fixed interval about
zero, $1/2\le\phi_\varepsilon\le3/2$ and $M_*/2\le|\phi_\varepsilon'|\le2M_*$
uniformly in $\varepsilon$. With
$\ell_\varepsilon=\phi_\varepsilon B_\varepsilon/(2\phi_\varepsilon')$ and
$\widetilde A_\varepsilon=\bigl(\begin{smallmatrix}A_\varepsilon&\ell_\varepsilon\\
\ell_\varepsilon^{\mathsf T}&1\end{smallmatrix}\bigr)$, the Schur complement
$1-\ell_\varepsilon^{\mathsf T}A_\varepsilon^{-1}\ell_\varepsilon$ is at least
$1/2$ for $M_*$ sufficiently large, and
$\partial_t(\phi_\varepsilon u)-\operatorname{tr}(\widetilde A_\varepsilon
D^2(\phi_\varepsilon u))=\phi_\varepsilon(u_t-\operatorname{tr}(A_\varepsilon D^2u)
-B_\varepsilon\cdot Du+c_\varepsilon u)$. This gives a uniformly elliptic lifting
with uniform smooth coefficient bounds, multiplication by $\phi_\varepsilon>0$
preserves time monotonicity, and for $\varepsilon=0$ one has
$\phi_0(\zeta)=1-M_*\zeta$. Thus the local obstacle compactness and the contact
cone regularity results remain applicable uniformly as the discount vanishes.

\begin{lemma}\label{lem:finite-axis-linear-lower}
For every compact interval $I\Subset(0,\infty)$ there are $c,C,y_0>0$
such that
\begin{equation}\label{eq:finite-axis-linear-lower}
 cy\le P(\tau,B(\tau),y)-K+B(\tau)\le Cy,
 \qquad \tau\in I,\quad 0<y<y_0.
\end{equation}
\end{lemma}
\begin{proof}
The upper bound is Proposition~\ref{prop:axis-linear-premium}. Suppose
the lower bound fails. Then there are $\tau_n\in I$ and
$\varepsilon_n\searrow0$ with
$\varepsilon_n^{-1}R(\tau_n,B(\tau_n),\varepsilon_n)\to0$, and after
passing to a subsequence $\tau_n\to\tau_0\in I$. Set
$x_n=\log B(\tau_n)$, $x_0=\log B(\tau_0)$, $f_0=rK-qB(\tau_0)>0$, and
\[
 W_n(t,X,Y)=\varepsilon_n^{-1}
 w(\tau_n+\varepsilon_n t,x_n+\varepsilon_n X,\varepsilon_n Y).
\]
For every compact subset of $\mathbb R\times\mathbb R\times(0,\infty)$,
these functions are uniformly bounded: the contact estimate
\eqref{eq:axis-w-bounds}, which is uniform for maturities in a
neighborhood of $I$, the stock Lipschitz estimate, and the time
Lipschitz estimate bound them by $C(Y+t^++X^+)$ after enlarging $C$
on the compact set. Their rescaled equations are
\[
 \min\{\partial_tW_n-\mathcal L_n
 W_n+f_n,W_n\}=0,
 \qquad f_n=rK-qB(\tau_n)e^{\varepsilon_n X},
\]
where
\begin{align*}
 \mathcal L_n={}&
 \frac Y2\partial_{XX}+\rho\sigma Y\partial_{XY}
 +\frac{\sigma^2Y}{2}\partial_{YY}
 +(\mu-\varepsilon_n Y/2)\partial_X
 +(a-\kappa\varepsilon_n Y)\partial_Y-\varepsilon_n r.
\end{align*}
On compact sets with $Y>0$, the coefficients and ellipticity constants
are uniform, and $f_n\to f_0$ locally uniformly by continuity of $B$.
The local obstacle estimates of Lemma~\ref{penalty}, followed by
compactness, give a subsequence converging locally uniformly, with
locally uniform convergence of the spatial first derivatives, to $W$.
Stock convexity gives
$W_{n,XX}-\varepsilon_n W_{n,X}\ge0$
in distributions, so convexity in $X$ passes to the limit.
The limit is nonnegative and nondecreasing in $t,X,Y$, convex in $X$,
and solves
\[
 \min\{\partial_tW-\mathcal L_0W+f_0,W\}=0,
 \quad
 \mathcal L_0=\frac Y2\partial_{XX}+\rho\sigma Y\partial_{XY}
 +\frac{\sigma^2Y}{2}\partial_{YY}+\mu\partial_X+a\partial_Y.
\]
By the contradiction hypothesis $W(0,0,1)=0$. Hence
\[
 W(t,X,Y)=0\quad\hbox{if }t\le0,\ X\le0,\ 0<Y<1.
\]
We next show that $W(0,X,Y)>0$ whenever $X>0$ and $Y>0$.
At $t=0$, every original point with $X>0$ belongs to continuation,
because $e^{x_n+\varepsilon_n X}>B(\tau_n)\ge b(\tau_n,\varepsilon_n Y)$.
Therefore time monotonicity and the continuation equation give
\[
 \mathcal L_n W_n(0,\cdot,\cdot)
 =f_n+\partial_tW_n(0,\cdot,\cdot)
 \ge f_n\quad\hbox{on }\{X>0,Y>0\}.
\]
This inequality passes to distributions in the spatial variables.
If $W(0,X_*,Y_*)=0$ with $X_*,Y_*>0$, monotonicity and
nonnegativity would make $W(0,\cdot,\cdot)$ identically zero on
$(0,X_*)\times(0,Y_*)$, contradicting $\mathcal L_0W(0,\cdot,\cdot)
\ge f_0>0$ there.

Fix $Y_*\in(0,1)$. The point $(0,0,Y_*)$ is a free-boundary point
of the limiting parabolic obstacle problem, and its spatial contact set
at time zero contains the half-ball $X\le0$. The limit problem has no
discount, so the uniform lifting described above replaces
Lemma~\ref{liftlemma}, and Proposition~\ref{contactcriterion} makes
$(0,0,Y_*)$ a regular free-boundary point.
Consequently $W$ is smooth from the continuation side near this point,
all its first derivatives have zero boundary trace, and its spatial
Hessian has rank one in the normal direction. At time zero the boundary
is precisely the flat line $X=0$ locally. Thus the tangential and mixed
second derivatives vanish, and the equation gives
\[
 W_{XX}(0,0+,Y_*)=\frac{2f_0}{Y_*}.
\]
For every fixed $Y\in(0,1)$ it follows that
\[
 W(0,X,Y)=\frac{f_0}{Y}X^2+o(X^2)\qquad(X\searrow0).
\]
Taking two fixed heights $0<Y_1<Y_2<1$ contradicts monotonicity of
$W$ in $Y$. This proves the lower bound.
\end{proof}

\begin{lemma}\label{lem:finite-axis-small-entry}
For every $M>0$ there are $p_M,\varepsilon_M>0$ such that the
process started at $(B,0)$, for any fixed $B>0$, satisfies
\[
 \mathbb P\left(0<Y_\varepsilon<\varepsilon,
 \ \log(S_\varepsilon/B)<-(M+1)\varepsilon\right)\ge p_M,
 \qquad 0<\varepsilon<\varepsilon_M.
\]
\end{lemma}
\begin{proof}
Brownian scaling identifies
$(\varepsilon^{-1}\log(S_{\varepsilon t}/B),
\varepsilon^{-1}Y_{\varepsilon t})$ on $0\le t\le1$ with
$(\Xi_t^\varepsilon,Z_t^\varepsilon)$ satisfying
\[
 dZ_t^\varepsilon=(a-\kappa\varepsilon Z_t^\varepsilon)dt
       +\sigma\sqrt{Z_t^\varepsilon}\,dW_t,
 \quad
 d\Xi_t^\varepsilon=(\mu-\varepsilon Z_t^\varepsilon/2)dt
       +\sqrt{Z_t^\varepsilon}\,dW^S_t,
 \quad Z_0^\varepsilon=\Xi_0^\varepsilon=0.
\]
Couple $Z^\varepsilon$ with $dZ_t=a\,dt+\sigma\sqrt{Z_t}\,dW_t$,
$Z_0=0$. Comparison and the explicit first moments give
$0\le Z^\varepsilon\le Z$ and
\[
 \mathbb E|Z_t-Z_t^\varepsilon|
 =at-\frac{a(1-e^{-\kappa\varepsilon t})}{\kappa\varepsilon}
 \le C\varepsilon t^2.
\]
Using $(\sqrt u-\sqrt v)^2\le|u-v|$ and It\^o isometry, we obtain
\[
 (\Xi_1^\varepsilon,Z_1^\varepsilon)
 \longrightarrow
 \left(\mu+\int_0^1\sqrt{Z_t}\,dW^S_t,Z_1\right)
 \quad\hbox{in probability}.
\]
The variable $Z_1$ has the gamma distribution of shape
$\nu=2a/\sigma^2$ and scale $\varsigma=\sigma^2/2$, as follows from
$\mathbb E e^{-\lambda Z_1}=(1+\varsigma\lambda)^{-\nu}$.
Conditionally on the variance Brownian motion $W$, the first
limiting coordinate is Gaussian with mean
$\mu+\rho(Z_1-a)/\sigma$ and variance
$(1-\rho^2)\int_0^1Z_t\,dt$. Its variance is positive on
$\{0<Z_1<1\}$, an event of positive probability.
The portmanteau inequality for open sets now proves
the assertion.
\end{proof}

\begin{theorem}\label{thm:finite-axis-separation}
For every $\tau>0$,
\[
 \lim_{y\searrow0}\frac{B(\tau)-b(\tau,y)}y=+\infty.
\]

\end{theorem}
\begin{proof}
Fix $\tau>0$ and abbreviate $B=B(\tau)$ and
$d(y)=\log(B/b(\tau,y))$. If the assertion failed, there would be
$M>0$ and $y_n\searrow0$ with $d(y_n)\le My_n$.
For all sufficiently large $n$, $y_n<\tau$, and on the event
\[
 E_n=\{0<Y_{y_n}<y_n,\
       \log(S_{y_n}/B)<-(M+1)y_n\}
\]
variance and maturity monotonicity imply
\[
 S_{y_n}<Be^{-My_n}\le b(\tau,y_n)
       \le b(\tau,Y_{y_n})\le b(\tau-y_n,Y_{y_n}).
\]
Thus the process is strictly inside the finite-maturity stopping region.
Lemma~\ref{lem:finite-axis-small-entry} yields $\liminf_n\mathbb P(E_n)>0$.
The event $\limsup_nE_n$ belongs to the completed germ sigma-field of
the driving Brownian motion, and the Blumenthal zero--one law
\cite[Chapter III, Theorem 2.15]{RY} makes its probability one.

Couple initial stocks by $S_t^{s,0}=sN_t$ and let $\eta_h\le\tau$
be the first contact time, including maturity, for the initial state
$(B+h,0)$. It is optimal by the finite-horizon Snell envelope theorem,
see \cite[Chapter I, Theorem 2.2]{PS} or \cite[Appendix D]{KS}.
Strict entry on the events $E_n$ implies $\eta_h\to0$ almost surely
as $h\searrow0$. Indeed, on $\limsup_nE_n$ there are, for every
$\varepsilon>0$, an index $n$ with $y_n<\varepsilon$ and a strict
inequality $S_{y_n}<b(\tau-y_n,Y_{y_n})$, which persists for the
initial stock $B+h$ when $h$ is small. Optimality and the stock
Lipschitz bound yield
\[
 -1\le\frac{P(\tau,B+h,0)-P(\tau,B,0)}h
 \le\mathbb E\left[e^{-r\eta_h}
 \frac{g((B+h)N_{\eta_h})-g(BN_{\eta_h})}h\right].
\]
The integrand is nonpositive and tends almost surely to $-1$.
Fatou's lemma applied to its negative proves
$P_s(\tau,B+,0)=-1$ under the contradiction hypothesis.
Convexity gives, for any fixed $\ell>0$,
\[
 0\le R_s(\tau,B,y)
 \le\frac{R(\tau,B+\ell,y)}\ell.
\]
Continuity of the value, followed by $\ell\searrow0$, proves
$R_s(\tau,B,y)\to0$ as $y\searrow0$.
The linear lower bound and convexity now imply along the chosen sequence
\[
 cy_n\le R(\tau,B,y_n)
 \le(B-b(\tau,y_n))R_s(\tau,B,y_n)
 \le BMy_nR_s(\tau,B,y_n)=o(y_n),
\]
a contradiction. Since $B-b\le BM y$ would itself imply
$d(y)=O(y)$ along a smaller sequence, the stated limit in the stock
coordinate follows.
\end{proof}
\begin{remark}\label{rem:axis-smooth-fit}
The proof shows that a boundary gap $O(y_n)$ along a sequence $y_n\searrow0$ at
some maturity would force smooth fit of the axis value function
$P(\tau,\cdot,0)$ at $B(\tau)$, and that smooth fit at the axis would in turn
force $R_s(\tau,B(\tau),y)\to0$ as $y\searrow0$. Neither statement is
established here, and whether $P(\tau,\cdot,0)$ is differentiable at $B(\tau)$
is left open. By convexity in the stock and joint continuity of $P$, this is
the same as asking whether $P_s$ extends continuously to the axis contact
point. The variance derivative does not extend continuously, see
Proposition~\ref{prop:axis-py-jump} below.
\end{remark}

\begin{corollary}\label{cor:axis-secant}
Fix $\tau>0$. For small $d>0$, let
$\gamma_\tau(d)$ denote the unique variance satisfying
$b(\tau,\gamma_\tau(d))=B(\tau)e^{-d}$. Then
\[
 \gamma_\tau(d)\longrightarrow0,
 \qquad\frac{\gamma_\tau(d)}d\longrightarrow0\quad(d\searrow0).
\]
Thus $(B(\tau)e^{-d},\gamma_\tau(d))$ has endpoint right derivative
$(-B(\tau),0)$. Continuity of the tangent is established below by rescaling at moving
contact points.
\end{corollary}
\begin{proof}
Endpoint matching and strict variance monotonicity give the inverse
and its continuity at zero. Substituting $y=\gamma_\tau(d)$ in
Theorem~\ref{thm:finite-axis-separation} and using
$1-e^{-d}\sim d$ gives $\gamma_\tau(d)/d\to0$.
\end{proof}

\subsection{The tangent of the slices and the variance derivative at the axis}
\label{subsec:axis-tangent}

We first record the stability input from the theory of regular points
used below.

\begin{lemma}\label{lem:normal-stability}
Let $u_k\ge0$ be time-monotone solutions of obstacle problems of the
form \eqref{generalobstacle} on a fixed cylinder, with operators whose
lifted versions are uniformly elliptic with uniformly bounded smooth
coefficients, with sources bounded away from zero uniformly, and with
uniformly bounded local solution norms. Suppose $u_k\to u$ locally
uniformly together with the first spatial derivatives, that $z_*$ is a
free-boundary point of every $u_k$ and of $u$ at time zero, and that
$z_*$ is a regular point of $u$ with inward spatial unit normal $e$.
Then the inward spatial unit normals $n_k$ of the free boundaries of
$u_k$ at $z_*$ converge to $e$.
\end{lemma}

\begin{proof}
Lift the equations with the fixed extra diagonal coefficient constructed
in this section, so that all operators are uniformly elliptic without
zeroth-order terms. At the regular limit point choose a sufficiently
small ordinary parabolic scale so that the rescaled limit is close in
$C^1$ to the half-space solution $q(Z)=\gamma_0(e\cdot Z)_+^2$.
For any $\lambda_0>0$ and every unit direction $v$ with $v\cdot e\ge\lambda_0$,
$\partial_vq-q\ge0$ on a sufficiently small common cylinder.
Spatial $C^1$ convergence transfers this inequality with an arbitrarily
small error to the approximating solutions. The estimate of the time
derivative in \cite[Proposition 6.1]{AK}, applied at their contact
centers, bounds the rescaled time derivative by an arbitrarily small
quantity when the scale is small. Hence, after the rescaling and a
normalization of the source, the lifted functions $u_k$ belong to the
class of \cite{AK} with uniform constants and satisfy, for $k$ large,
the hypothesis $C\partial_vu_k-\partial_tu_k-u_k\ge-\varepsilon_0$ of
\cite[Lemma 8.2]{AK} on the unit cylinder, with the small constant
$\varepsilon_0$ required there. All constants are uniform because
the lifted ellipticity, coefficient bounds, source bounds, and local
solution bounds are uniform. Lemma 8.2 of \cite{AK} then makes
$C\partial_v u_k-\partial_tu_k-u_k$ nonnegative on a smaller cylinder.
Since $\partial_tu_k\ge0$, each such $v$ is a spatial direction of
monotonicity of $u_k$, so the contact set of $u_k$ at time zero is
invariant under small translations in the direction $-v$. At the smooth
contact point $z_*$ this implies $v\cdot n_k\ge0$. The cone of
directions $v\cdot e\ge\lambda_0$ forces $n_k$ into a cone about $e$ whose
opening tends to zero with $\lambda_0$. This proves normal convergence. This
is the almost positivity argument behind
\cite[Propositions 8.3 and 8.5]{AK}. For the linear lifted operators,
the bounded-Hessian interpretation of their coefficient condition is
as in Remark~\ref{AKremark}.
\end{proof}

\begin{proposition}\label{prop:finite-axis-normals}
For every fixed $\tau>0$,
\[
 b_y(\tau,y)\longrightarrow-\infty\qquad(y\searrow0).
\]
The closure of the positive-variance boundary slice at $(B(\tau),0)$
is a regular $C^1$ curve tangent to $y=0$.
\end{proposition}
\begin{proof}
Fix $\tau$ and set $\beta_\varepsilon=\log b(\tau,\varepsilon)$,
$\beta_0=\log B(\tau)$, and $d_\varepsilon=\beta_0-\beta_\varepsilon$.
Theorem~\ref{thm:axis-cross-maturity} gives $d_\varepsilon\to0$ and
Theorem~\ref{thm:finite-axis-separation} gives
$d_\varepsilon/\varepsilon\to\infty$. Consider
\[
 F_\varepsilon(t,X,Y)=\varepsilon^{-1}
 w(\tau+\varepsilon t,\beta_\varepsilon+\varepsilon X,
      \varepsilon Y).
\]
On every fixed compact set, the stock argument is below $B(\tau)$
for small $\varepsilon$. The contact premium estimate and the maturity
Lipschitz estimate give
\[
 0\le F_\varepsilon(t,X,Y)\le C(Y+|t|).
\]
Convexity in stock and the secant inequality at $B(\tau)$ also give
\[
 0\le\partial_XF_\varepsilon(t,X,Y)
 \le \frac{C\varepsilon(Y+|t|)}{d_\varepsilon}
 \longrightarrow0
\]
uniformly on every fixed compact set. As before, uniform lifting
and local obstacle compactness produce a subsequential limit $F$,
nondecreasing in time and in $Y$, and independent of $X$.
The point $z_*=(0,0,1)$ is a free-boundary point of every rescaled
problem. At time zero, the quadrant $X\le0$, $Y\le1$ belongs to
the contact set. Hence $F(0,X,Y)=0$ for every $X$ and $Y\le1$.
At time zero the continuation equation gives
$\mathcal L_\varepsilon F_\varepsilon\ge f_*:=(rK-qB(\tau))/2>0$ for
small $\varepsilon$. The usual
quadratic barrier on a fixed spatial ball, using the elliptic maximum
principle in its continuation part, gives
$\sup_{B_r(z_*)}F_\varepsilon(0,\cdot)\ge c r^2$ with uniform
$c>0$. Thus spatial nondegeneracy ensures that
$z_*$ remains a free-boundary point of $F$. It cannot become an
interior contact point. The contact half-space and
Proposition~\ref{contactcriterion} make $z_*$ a regular point of $F$.
Since $F$ is independent of $X$, its inward spatial unit normal there
is $e_Y$. Lemma~\ref{lem:normal-stability} now gives convergence to
$e_Y$ of the normals of $F_\varepsilon$ at $z_*$. These normals are
\[
 n_\varepsilon=
 \frac{(1,-\beta_y(\tau,\varepsilon))}
 {\sqrt{1+\beta_y(\tau,\varepsilon)^2}}.
\]
Thus $\beta_y(\tau,\varepsilon)\to-\infty$, and
$b_y=b\beta_y\to-\infty$, since
$b(\tau,\varepsilon)\to B(\tau)>0$.
Every subsequence has the same conclusion, proving the full limit.

Strict variance monotonicity supplies a smooth inverse $y=\Gamma_\tau(s)$
for $s<B(\tau)$ sufficiently close to $B(\tau)$.
Its continuous extension satisfies $\Gamma_\tau(B(\tau))=0$ and
\[
 \Gamma_\tau'(s)=\frac{1}{b_y(\tau,\Gamma_\tau(s))}
 \longrightarrow0\qquad(s\nearrow B(\tau)).
\]
The mean value theorem then gives differentiability of the extension
at the endpoint with derivative zero. The parametrization
$s\mapsto(s,\Gamma_\tau(s))$ has nonvanishing derivative and
horizontal endpoint tangent, which proves the geometric statement.
\end{proof}

The inverse graphs of the slices fit together into a single function.
By strict variance monotonicity, the free boundary near the axis is the
graph
\begin{equation}\label{eq:axis-Gamma}
 y=\Gamma(\tau,s),\qquad 0<\tau,\quad b(\tau,1)<s<B(\tau),
\end{equation}
where $\Gamma(\tau,\cdot)=\Gamma_\tau$ takes values in $(0,1)$, and
$\Gamma$ is smooth on the open set where it is defined by the implicit
function theorem. The
function $\Gamma$ extends continuously by $\Gamma(\tau,B(\tau))=0$,
because $b$ is jointly continuous at the axis.

\begin{proposition}\label{prop:axis-Gamma-partials}
Let $\tau>0$. Both partial derivatives of the inverse graph
\eqref{eq:axis-Gamma} vanish at the axis along the maturity slice,
\[
 \Gamma_s(\tau,s)\longrightarrow0,\qquad
 \Gamma_\tau(\tau,s)\longrightarrow0\qquad(s\nearrow B(\tau)).
\]
Moreover, for every compact interval $I\Subset(0,\infty)$ there is a
constant $C_I$ such that
\begin{equation}\label{eq:axis-Gamma-ratio}
 \abs{\Gamma_\tau(\tau,s)}\le C_I\,\abs{\Gamma_s(\tau,s)}
\end{equation}
on the part of the domain of $\Gamma$ with $\tau\in I$ and
$\Gamma(\tau,s)\le1$.
\end{proposition}
\begin{proof}
Differentiating $b(\tau,\Gamma(\tau,s))=s$ gives
\[
 \Gamma_s=\frac1{b_y(\tau,\Gamma)},\qquad
 \Gamma_\tau=-\frac{b_\tau(\tau,\Gamma)}{b_y(\tau,\Gamma)}.
\]
Estimate \eqref{eq:axis-btau-bound} bounds $\abs{b_\tau}$ by
$C_I(1+Y_*)$ for $\tau\in I$ and $0<y\le Y_*$, which gives
\eqref{eq:axis-Gamma-ratio} with $Y_*=1$. At a fixed maturity,
$\Gamma(\tau,s)\to0$ as $s\nearrow B(\tau)$, and
Proposition~\ref{prop:finite-axis-normals} gives $b_y\to-\infty$ along
the slice. Both limits follow.
\end{proof}

\begin{remark}\label{rem:axis-C1-surface}
Proposition~\ref{prop:axis-Gamma-partials} is the slice form of $C^1$
regularity of the closed free boundary up to its edge on the axis. The full
statement would be that $\Gamma$, extended by zero across the Lipschitz curve
$s=B(\tau)$, is continuously differentiable on a neighborhood of that curve
with vanishing gradient on it. By \eqref{eq:axis-Gamma-ratio}, and by
integrating $-\Gamma_s$ along horizontal segments ending on the curve, this
holds as soon as $\Gamma_s(\tau',s')\to0$ as $(\tau',s')\to(\tau,B(\tau))$
within the domain of $\Gamma$. Since $\Gamma_s=1/b_y$, the missing statement
is the joint limit $b_y(\tau',y)\to-\infty$ as $(\tau',y)\to(\tau,0)$, and
inspection of the proof of Proposition~\ref{prop:finite-axis-normals} shows
that it follows once the separation of
Theorem~\ref{thm:finite-axis-separation} holds locally uniformly in $\tau$.
The proof of that theorem uses the Blumenthal law along a sequence of
variances at one fixed maturity, and we have not been able to make it
uniform. Should the joint limit hold, the closure of the interior free
boundary would be a $C^1$ surface with boundary whose tangent plane along
the axis is the plane $y=0$.
\end{remark}

On the contact set the price equals the payoff, so $P_y=0$ there. The
linear nondegeneracy of Lemma~\ref{lem:finite-axis-linear-lower} shows
that this value is not the limit of $P_y$ along the axis contact stock.

\begin{proposition}\label{prop:axis-py-jump}
Let $I\Subset(0,\infty)$ be compact and let $c$ be the constant of
\eqref{eq:finite-axis-linear-lower}. For every $\tau\in I$,
\[
 \limsup_{y\searrow0}P_y(\tau,B(\tau),y)\ge c>0,
 \qquad
 \liminf_{y\searrow0}\frac{P(\tau,B(\tau),y)-K+B(\tau)}{y}\ge c.
\]
Consequently the variance derivative $P_y$, which vanishes identically
on the contact set, does not extend continuously to the point
$(\tau,B(\tau),0)$, and $P(\tau,\cdot,\cdot)$ is not continuously
differentiable on any neighborhood of $(B(\tau),0)$ in the closed
half-plane $y\ge0$. The same conclusion holds along every stock
$s\in(B(\tau)-cy/2,B(\tau)]$ at height $y$, in the sense that
\[
 R(\tau,s,y)\ge\frac{cy}2\qquad
 (\tau\in I,\ 0<y<y_0,\ B(\tau)-cy/2\le s\le B(\tau)).
\]
\end{proposition}
\begin{proof}
For $y>0$ the point $(\tau,B(\tau),y)$ lies in the continuation region,
since $b(\tau,y)<B(\tau)$ by Theorem~\ref{thm:finite-axis-separation},
and $P$ is smooth there with $P_y\ge0$. For $0<\delta<y$ the
fundamental theorem of calculus gives
$R(\tau,B(\tau),y)-R(\tau,B(\tau),\delta)
=\int_\delta^yP_y(\tau,B(\tau),v)\,dv$, and letting $\delta\searrow0$
with $R(\tau,B(\tau),0)=0$ and monotone convergence yields
\[
 R(\tau,B(\tau),y)=\int_0^yP_y(\tau,B(\tau),v)\,dv
 \le y\sup_{0<v\le y}P_y(\tau,B(\tau),v).
\]
The lower bound $R(\tau,B(\tau),y)\ge cy$ therefore forces
$\sup_{0<v\le y}P_y(\tau,B(\tau),v)\ge c$ for every small $y$, which is
the first assertion. The second is \eqref{eq:finite-axis-linear-lower}.
Every contact point $(\tau,s,y')$ with $s<B(\tau)$ and $0<y'\le
\Gamma_\tau(s)$ has $P_y=0$, and these points accumulate at
$(\tau,B(\tau),0)$. Hence $P_y$ has no continuous extension there.
The final estimate follows from the $1$-Lipschitz property of $R$ in the
stock, $R(\tau,s,y)\ge R(\tau,B(\tau),y)-(B(\tau)-s)\ge cy/2$.
\end{proof}

\begin{remark}\label{rem:axis-py-geometry}
Proposition~\ref{prop:axis-py-jump} quantifies how thin the contact set is
near the axis. Every stock within $cy/2$ below $B(\tau)$ carries a premium of
at least $cy/2$ at height $y$, so $B(\tau)-b(\tau,y)\ge cy/2$ uniformly on
compact maturity intervals, and the variance derivative, which vanishes on
the contact set, is at least $c$ at points $(\tau,B(\tau),y_n)$ with
$y_n\searrow0$. Smooth fit thus fails in the degenerate direction at the
axis, while it holds in every direction at positive variance by
Theorem~\ref{main}. No upper bound for $P_y$ near the axis contact point is
claimed.
\end{remark}

The inverse tangency determines the limiting boundary Hessian even
though it does not determine a sharp equivalent for the boundary gap.

\begin{corollary}\label{cor:axis-weighted-limit}
For each $\tau>0$, with the derivatives evaluated on the continuation
side of $x=\beta(\tau,y)$,
\begin{equation}\label{eq:axis-weighted-limit}
 \lim_{y\searrow0} yD^2_{x,y}w^+
 =\lim_{y\searrow0} yD^2_{x,y}U^+
 =\frac{2(rK-qB(\tau))}{\sigma^2}
       \begin{pmatrix}0&0\\0&1\end{pmatrix}.
\end{equation}
In stock coordinates this implies
\[
 yP_{ss}(\tau,b(\tau,y)+,y)\longrightarrow0,\qquad
 yP_{sy}(\tau,b(\tau,y)+,y)\longrightarrow0,
\]
whereas
$yP_{yy}(\tau,b(\tau,y)+,y)\to2(rK-qB(\tau))/\sigma^2$.
These statements concern boundary traces from the continuation side only.
\end{corollary}
\begin{proof}
By Proposition~\ref{prop:finite-axis-normals},
$\beta_y=b_y/b\to-\infty$. Divide the numerator matrix and the
denominator in \eqref{hessianmain} by $\beta_y^2$ and use
$b(\tau,y)\to B(\tau)$ to obtain the first limit.
Since $U=w+K-e^x$, multiplication by $y$ removes the bounded Hessian
of $K-e^x$, which proves the second.
At the boundary $P_s=-1$, so
$w_{xx}=b^2P_{ss}$, $w_{xy}=bP_{sy}$, and $w_{yy}=P_{yy}$.
Finally $b\to B(\tau)>0$, giving all three limits in stock coordinates.
\end{proof}

The proof of Theorem~\ref{mainaxis} is completed in
Subsection~\ref{subsec:expiry}, after the expiry limit of the boundary
has been established.

\subsection{Exact vertical identities and the maturity contribution}
\label{subsec:finite-vertical}

Fix $\tau>0$ and write $\beta_0(\tau)=\log B(\tau)$. Strict variance
monotonicity, interior smoothness and endpoint matching show that, for
$x$ in a sufficiently short interval immediately below $\beta_0(\tau)$,
there is a unique $\ell_\tau(x)>0$ with $\beta(\tau,\ell_\tau(x))=x$, namely
$\ell_\tau(x)=\Gamma_\tau(e^x)$ in the notation of
Proposition~\ref{prop:finite-axis-normals}. Moreover $\ell_\tau$ is
smooth, $\ell_\tau'(x)<0$, and $\ell_\tau(x)\to0$ as $x\nearrow\beta_0(\tau)$.
The contact interval on this vertical line is $0\le y\le \ell_\tau(x)$.

Define the spatial remainder, extended by zero in contact, by
\begin{equation}\label{eq:finite-E}
 \mathcal E(\tau,x,y)=\kappa y w_y-\frac y2w_{xx}
 -\rho\sigma y w_{xy}-(r-q-y/2)w_x+rw.
\end{equation}
For $0<t<h$ and $0<\ell\le h$, put
\begin{align}
 K_\nu(h,t)&=\frac{t^{\nu-1}}{\varsigma}\int_t^h z^{-\nu}\,dz,
 \label{eq:finite-kernel}\\
 \Psi_\nu(\ell,h)&=\ell+\ell^\nu\int_\ell^h z^{-\nu}\,dz.
 \label{eq:finite-Psi}
\end{align}
Set $\Psi_\nu(0,h)=0$ and $\ell_h=\min\{\ell_\tau(x),h\}$.

\begin{proposition}\label{prop:finite-vertical}
For every such $(\tau,x)$ and every sufficiently small $h>0$,
\begin{align}
 \frac{f(x)}a \ell_h^\nu
 &=h^\nu\left(\frac{f(x)}a-w_y(\tau,x,h)\right)
   +\frac1{\varsigma}\int_0^h t^{\nu-1}(\mathcal E+w_\tau)(\tau,x,t)\,dt,
 \label{eq:finite-flux}\\
 w(\tau,x,h)
 &=\frac{f(x)}a\bigl[h-\Psi_\nu(\ell_h,h)\bigr]
   +\int_0^hK_\nu(h,t)\mathcal E(\tau,x,t)\,dt
   +\mathcal T_h(\tau,x),\label{eq:finite-value}\\
 \mathcal T_h(\tau,x)
 &=\int_0^h K_\nu(h,t)w_\tau(\tau,x,t)\,dt\ge0.
 \label{eq:finite-time-remainder}
\end{align}
The identities also hold distributionally in $x$ on compact
subintervals below $\beta_0(\tau)$.
\end{proposition}

\begin{proof}
The interior result already proved gives smoothness on each side of
the free boundary, zero gradient at contact, and locally bounded second
spatial derivatives across it.  On a compact interval of $x$ under
consideration, the contact height is bounded below by a positive
constant.  Thus every integrand vanishes near $y=0$, and no derivative
trace at the degenerate axis is being assumed.  The obstacle equation
can be rearranged as
\[
 \varsigma yw_{yy}+aw_y
 =f(x)\mathbf1_{\{y>\ell_\tau(x)\}}+\mathcal E+w_\tau.
\]
Since $a=\varsigma\nu$, multiplication by $y^{\nu-1}$ makes the left-hand
side $\varsigma\partial_y(y^\nu w_y)$.  Integration from the lower contact
interval to $h$, where $w_y=0$ at the lower endpoint, proves
\eqref{eq:finite-flux}.  A second integration, with $w=0$ at the
lower endpoint, and Fubini's theorem give \eqref{eq:finite-value}.
The contribution of the constant source is
\[
 \frac{f(x)}a\int_{\ell_h}^h[1-(\ell_h/t)^\nu]\,dt
 =\frac{f(x)}a[h-\Psi_\nu(\ell_h,h)].
\]
The case $h\le \ell_\tau(x)$ gives zero on both sides.  Finally,
$K_\nu\ge0$ and $w_\tau\ge0$ prove the sign in
\eqref{eq:finite-time-remainder}.  Integrating the same calculation
against a compactly supported smooth function proves the weak identity.
\end{proof}

\begin{figure}[t]
 \centering
 \includegraphics[width=0.88\textwidth]{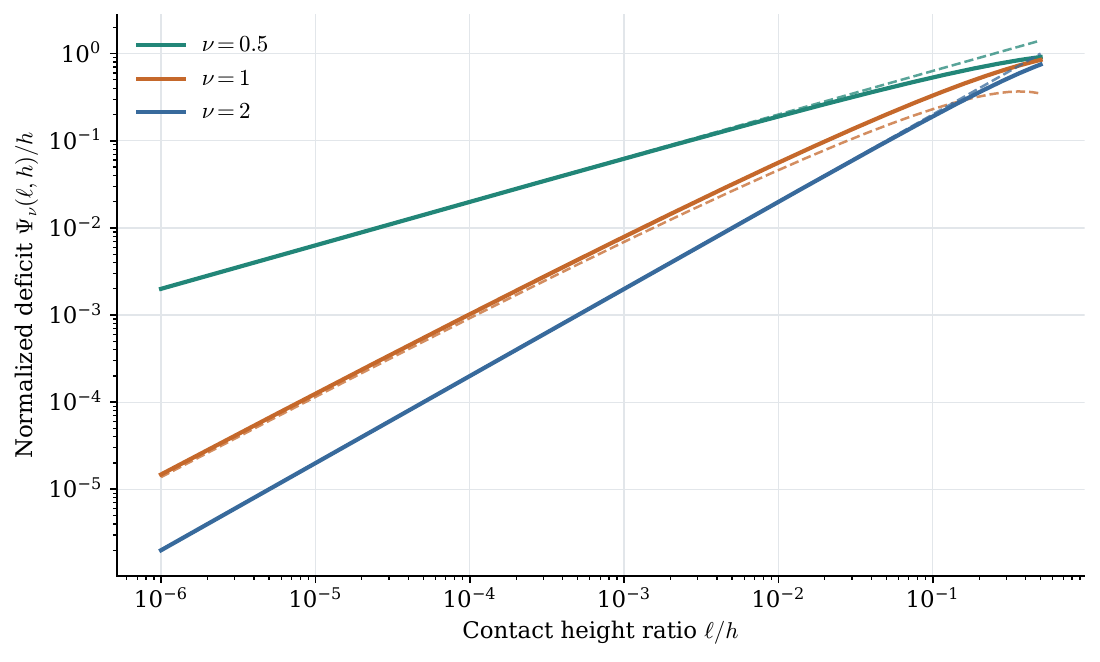}
 \caption{The exact normalized vertical deficit
 $\Psi_\nu(\ell,h)/h$ as a function of $\ell/h$, together with its
 leading expression for small ratios (dashed), for three values of the CIR
 index.  These curves evaluate \eqref{eq:finite-Psi-explicit}. They
 are not fitted exercise boundaries or numerical evidence for a sharp
 finite-maturity boundary law.}
 \label{fig:vertical-deficit}
\end{figure}

The kernel deficit has the explicit form
\begin{equation}\label{eq:finite-Psi-explicit}
 \Psi_\nu(\ell,h)=
 \begin{cases}
 (\ell^\nu h^{1-\nu}-\nu \ell)/(1-\nu),&\nu\ne1,\\
 \ell[1+\log(h/\ell)],&\nu=1.
 \end{cases}
\end{equation}
In particular, as $\ell/h\searrow0$,
\begin{equation}\label{eq:finite-Psi-rates}
 \Psi_\nu(\ell,h)\sim
 \begin{cases}
 \ell^\nu h^{1-\nu}/(1-\nu),&0<\nu<1,\\
 \ell\log(h/\ell),&\nu=1,\\
 \nu \ell/(\nu-1),&\nu>1.
 \end{cases}
\end{equation}
These are exact kernel asymptotics, plotted in
Figure~\ref{fig:vertical-deficit}.  They do not assert asymptotic
equivalents for $B(\tau)-b(\tau,y)$.

The time contribution has a useful integrated bound.  Let $J$ be a
compact time interval and let $\chi\ge0$ be smooth and supported in a
logarithmic stock interval which lies below $\beta_0(\tau)$ for every
$\tau\in J$.  For $\psi\in C_c^\infty(J)$, set
$A_h(\tau,x)=\int_0^hK_\nu(h,t)w(\tau,x,t)\,dt$.
The linear premium estimate gives
\[
 |A_h(\tau,x)|\le Ch^2,
 \qquad \int_0^h tK_\nu(h,t)\,dt
           =\frac{h^2}{2\varsigma(\nu+1)}.
\]
Since $\mathcal T_h=\partial_\tau A_h$ in distributions,
\begin{equation}\label{eq:finite-time-averaged}
 \left|\int_J\!\int\psi(\tau)\chi(x)\mathcal T_h(\tau,x)
                              \,dx\,d\tau\right|
 \le Ch^2\|\psi'\|_{L^1}\|\chi\|_{L^1}.
\end{equation}
The pointwise estimate provided by Lemma~\ref{lem:axis-time-lipschitz} is only
$\mathcal T_h\le Ch$, because $\int_0^hK_\nu(h,t)\,dt=h/a$.
Thus a small averaged remainder does not justify discarding the
maturity term at a fixed time.

\subsection{The boundary at expiry}
\label{subsec:expiry}

The results so far concern positive remaining maturities. At expiry the
payoff has a kink at the strike, and Section~\ref{sec:geometry} gives
no information on the boundary as $\tau\searrow0$. The next result
shows that the whole boundary surface, including its axis edge,
converges to the strike at expiry. The argument is the finite-maturity
analogue of the classical fact that the Black and Scholes put boundary
tends to $K$ at expiry when $q\le r$, see \cite{Kim,Jacka,CC}, and it
uses only the parabolic equation in the continuation region, the
continuity of the price at expiry and its monotonicity in remaining
maturity.

\begin{proposition}\label{prop:expiry}
For every $y\ge0$, with $b(\tau,0)=B(\tau)$,
\[
 \lim_{\tau\searrow0}b(\tau,y)=K .
\]
The convergence is uniform on bounded variance intervals $[0,Y_*]$. In
particular $b$ extends to a continuous function on $[0,\infty)^2$ with
$b(0,y)=K$, and $B(0+)=K$.
\end{proposition}

\begin{proof}
Fix $y_1>0$ and put $b_0=\lim_{\tau\searrow0}b(\tau,y_1)$, which
exists by maturity monotonicity and equals $\sup_{\tau>0}b(\tau,y_1)$.
Suppose $b_0<K$. Choose $b_0<s_1<s_2<K$ and $y_1<y_2$, and let
$\Omega=(\log s_1,\log s_2)\times(y_1,y_2)$. For every $\tau>0$ and
$(x,y)\in\Omega$ one has $e^x>b_0\ge b(\tau,y_1)\ge b(\tau,y)$, so the
whole cylinder $(0,\infty)\times\Omega$ lies in the continuation
region. There the function $U$ of \eqref{logdefs} is smooth and solves
$\op U=0$. Moreover $U(\tau,\cdot)\to K-e^x$ uniformly on
$\overline\Omega$ as $\tau\searrow0$, because $e^x<K$ on
$\overline\Omega$ and, for every $s>0$ and $y\ge0$,
\begin{align*}
 0\le P(\tau,s,y)-(K-s)^+
 &\le\sup_{\eta\le\tau}\E_{s,y}\abs{e^{-r\eta}(K-S_\eta)^+-(K-s)^+}\\
 &\le K(1-e^{-r\tau})+s\,\E_y\sup_{u\le\tau}\abs{N^y_u-1},
\end{align*}
where the last expectation tends to zero as $\tau\searrow0$ by
continuity of paths and dominated convergence, uniformly for $y$ in
compact sets. The logarithmic gap $w=U-K+e^x$ is therefore nonnegative
on $(0,\infty)\times\overline\Omega$, tends to zero uniformly on
$\overline\Omega$ as $\tau\searrow0$, and satisfies $\op w=-f$ on
$(0,\infty)\times\Omega$ by \eqref{forcing}, where
$f\ge f_1:=rK-qs_2>0$ on $\Omega$. Maturity monotonicity of the price
gives $w_\tau\ge0$ on $(0,\infty)\times\Omega$.

Let $\varphi\in C^\infty_c(\Omega)$ be nonnegative and not identically
zero. For $\tau>0$ the function $w$ is smooth on a neighborhood of
$\supp\varphi$, and the equation gives
\[
 0\le\int_\Omega w_\tau(\tau,\cdot)\varphi
 =\int_\Omega w(\tau,\cdot)\,(\Lh^*\varphi-r\varphi)-\int_\Omega f\varphi,
\]
where $\Lh^*$ is the formal adjoint of $\Lh$ and the integration by
parts produces no boundary terms because $\varphi$ has compact support
in $\Omega$. As $\tau\searrow0$ the first integral on the right tends
to zero, since $w(\tau,\cdot)\to0$ uniformly on $\supp\varphi$, whereas
the second is at least $f_1\int_\Omega\varphi>0$. For small $\tau$ the
right-hand side is therefore negative, which contradicts the left
inequality. Therefore $b_0=K$.

For $0\le y\le y_1$, variance monotonicity of the contact set gives
$b(\tau,y_1)\le b(\tau,y)\le K$, where at $y=0$ the inequality
$B(\tau)\ge b(\tau,y_1)$ holds because $P(\tau,s,0)\le P(\tau,s,y_1)$.
Since $b(\tau,y_1)\to K$, the convergence is uniform on $[0,y_1]$, and
$y_1$ is arbitrary. Joint continuity on $[0,\infty)^2$ follows from
this uniform convergence together with the continuity on
$(0,\infty)\times[0,\infty)$ established above.
\end{proof}

\begin{remark}\label{rem:expiry-corner}
Proposition~\ref{prop:expiry} settles the value of the boundary at the corner
$\tau=0$, $y=0$, but not its shape. In one dimension the boundary approaches
the strike with a logarithmic correction to the square-root law, see
\cite{CC}, and along the axis the integrated variance over a short time
$\tau$ is of order $\tau^2$, so a different scale is expected there. Rates at
the corner are not attempted here.
\end{remark}

\begin{proof}[Proof of Theorem~\ref{mainaxis}]
Part~(i) is the inequality $0<B(\tau)<K$ from the opening paragraph
of this section, Proposition~\ref{prop:axis-maturity-lipschitz}, the
matching identity and joint continuity of
Theorem~\ref{thm:axis-cross-maturity}, and Proposition~\ref{prop:expiry}.
Part~(ii) is Lemma~\ref{lem:axis-time-lipschitz},
Proposition~\ref{prop:axis-linear-premium} and
Lemma~\ref{lem:finite-axis-linear-lower}. Part~(iii) is
Theorem~\ref{thm:finite-axis-separation},
Proposition~\ref{prop:finite-axis-normals} and
Proposition~\ref{prop:axis-Gamma-partials}. Part~(iv) is
Theorem~\ref{thm:axis-quantitative-modulus}. Part~(v) is
Corollary~\ref{cor:axis-weighted-limit}, and part~(vi) is
Proposition~\ref{prop:axis-py-jump}.
\end{proof}

\subsection{The moving endpoint and sharp boundary equivalents}
\label{subsec:moving-endpoint}

There is an additional transport term when the zero-variance contact
point moves with remaining maturity. The local Lipschitz estimate gives
$\beta_0=\log B\in W^{1,\infty}_{\mathrm{loc}}((0,\infty))$,
so its derivative exists almost everywhere and is bounded on compact
intervals. Set
\[
 \widehat w(\tau,z,y)=w(\tau,\beta_0(\tau)+z,y).
\]
In continuation, the Sobolev chain rule gives, for almost every maturity,
\begin{equation}\label{eq:moving-axis-equation}
 \begin{split}
 \widehat w_\tau={}&\frac y2
 (\widehat w_{zz}+2\rho\sigma\widehat w_{zy}
                          +\sigma^2\widehat w_{yy})
 +(r-q+\beta_0'(\tau)-y/2)\widehat w_z\\
 &+(a-\kappa y)\widehat w_y-r\widehat w
       -f(\beta_0(\tau)+z).
 \end{split}
\end{equation}
The identity also holds distributionally in time. If one freezes a time
at which $\beta_0$ is differentiable
and removes the mixed derivative, the spatial drift parameters are
therefore
\begin{equation}\label{eq:finite-effective-drift}
 \mu_\mathrm{eff}(\tau)=r-q+\partial_\tau\log B(\tau),\qquad
 c_\mathrm{eff}(\tau)=
 \frac{2\mu_\mathrm{eff}(\tau)/\sigma-\rho\nu}{\sqrt{1-\rho^2}}.
\end{equation}
The sign of the endpoint speed term is tied to remaining maturity. In
calendar time it is $-\partial_t\log B$. By
Proposition~\ref{prop:axis-maturity-lipschitz} the correction
$\partial_\tau\log B$ is nonpositive and bounded on compact maturity
intervals, so $\mu_\mathrm{eff}$ ranges in a compact subinterval of
$(-\infty,r-q]$.

The intrinsic scaling at $(\tau_0,x_0,0)$ is
\[
 W_\varepsilon(T,X,Y)=\varepsilon^{-1}
 w(\tau_0+\varepsilon T,x_0+\varepsilon X,\varepsilon Y).
\]
Its continuation equation has a nonvanishing $\partial_TW_\varepsilon$
term.  The limiting equation is parabolic, with time and variance
both scaled by $\varepsilon$.  Freezing time is consequently an
additional analytical step, not a consequence of the smallness of
variance.

We close with a list of what the results of this section leave open.
The endpoint is locally Lipschitz, the boundary moves with bounded
speed in maturity up to the axis, and its spatial slices have regular
$C^1$ closures whose inverse graphs have vanishing partial derivatives
in both variables along each slice. These results do not establish
differentiability of $B$ at every positive maturity, the joint
vanishing of the gradient of the inverse graph at the axis discussed in
Remark~\ref{rem:axis-C1-surface}, the differentiability of the axis
value $P(\tau,\cdot,0)$ at $B(\tau)$ discussed in
Remark~\ref{rem:axis-smooth-fit}, or the decay and tangential
regularity of $\widehat w_\tau$ required to replace the parabolic
problem by the stationary axis problem. In particular, no sharp
logarithmic, critical exponential, or fractional-power equivalent for
the finite-maturity boundary gap is claimed. A complete derivation of
such equivalents requires stronger temporal endpoint regularity and
control of the time-dependent kernel contribution in
\eqref{eq:finite-time-remainder}. Formula
\eqref{eq:finite-effective-drift} identifies the correction that any
such derivation must retain.

\section{Numerical illustration}\label{sec:numerics}

We illustrate the interior boundary surface with a standard finite difference
scheme in the variables $(x,y)$ with $x=\log(s/K)$, for the two parameter
sets of Table~\ref{tab:params}. The linear
complementarity problem for $U$ was discretized on a uniform grid with
central differences for the pure second derivatives, the seven-point stencil
of Ikonen and Toivanen \cite{IT} for the mixed derivative, which keeps the
off-diagonal entries of the discrete generator nonnegative when the grid
ratio satisfies $\abs\rho\sigma\,\Delta x\le\Delta y\le\sigma\Delta x/\abs\rho$,
and central differences for the first derivatives, replaced by upwind
differences at the few nodes near $y=0$ where the central choice would
violate the sign condition. On the line $y=0$ the equation degenerates to a
first-order transport equation and was discretized with forward differences,
so no boundary condition is imposed there beyond the equation itself. The
early-exercise constraint was handled by the operator splitting method of
\cite{IT} with implicit Euler time stepping, and the Dirichlet values
$U=K-s$ at the lower end of the log-price grid and $U=0$ at the upper end,
together with a homogeneous Neumann condition at the upper variance level,
close the system. The location of the boundary at each variance level was
extracted from the quadratic contact behavior \eqref{taylor}, since
$\sqrt{w}$ is asymptotically linear in $x-\beta$ and its zero can be
extrapolated from the first two grid nodes on the continuation side.

\begin{table}[ht]
\centering
\caption{Parameters of the two examples. Set A satisfies
$2\kappa\theta\ge\sigma^2$ and set B violates it. In both cases $K=1$,
$r=0.05$, $q=0.02$, $\kappa=2$, $\theta=0.04$, $\rho=-0.7$, $T=1$.}
\label{tab:params}
\begin{tabular}{@{}lcccc@{}}
\toprule
 & $\sigma$ & $2\kappa\theta$ & grid $(x,y,\tau)$ & $\Delta x,\ \Delta y,\ \Delta\tau$\\
\midrule
Set A & $0.3$ & $0.16>0.09$ & $705\times601\times2000$ & $0.0062,\ 0.002,\ 0.0005$\\
Set B & $0.6$ & $0.16<0.36$ & $705\times301\times2000$ & $0.0062,\ 0.004,\ 0.0005$\\
\bottomrule
\end{tabular}
\end{table}

\begin{figure}[htbp]
\centering
\includegraphics[width=0.66\textwidth]{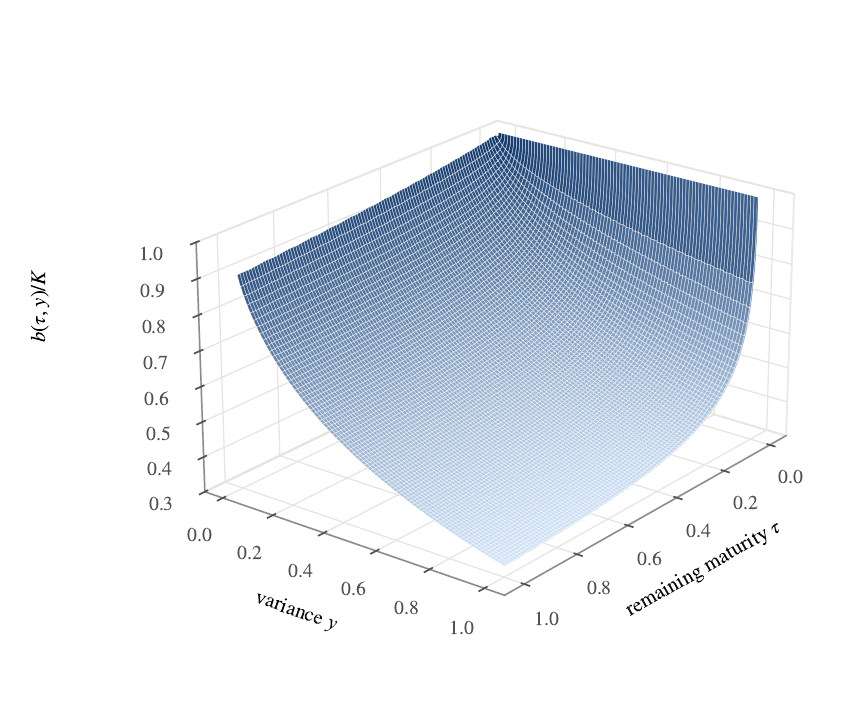}
\caption{The computed exercise boundary $b(\tau,y)/K$ of the finite-maturity
Heston American put for parameter set A. The surface is decreasing in both
remaining maturity $\tau$ and variance $y$, and it is visibly smooth away
from $\tau=0$, in agreement with Theorem~\ref{main}. The ridge at $\tau=0$ is
the kink of the payoff at the strike.}
\label{fig:surface}
\end{figure}

\begin{figure}[htbp]
\centering
\includegraphics[width=0.92\textwidth]{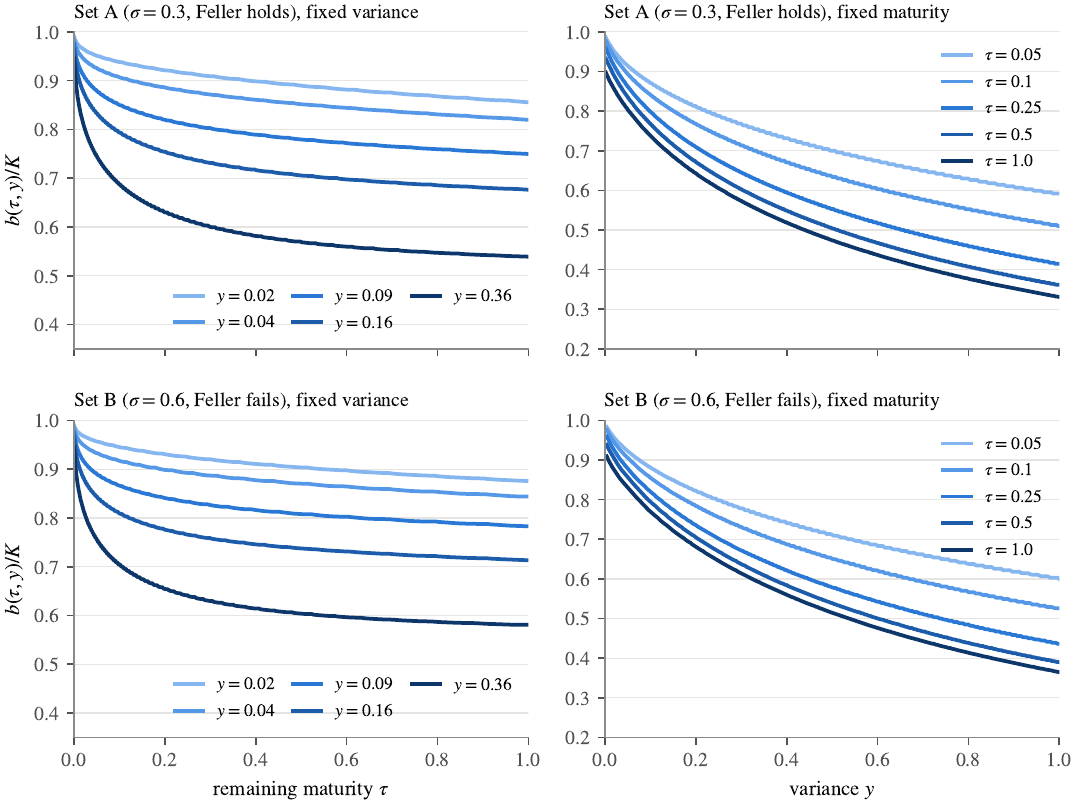}
\caption{Sections of the computed exercise boundary. Left column, $b/K$ as
a function of remaining maturity at fixed variance levels. Right column,
$b/K$ as a function of variance at fixed remaining maturities. Top row,
parameter set A. Bottom row, parameter set B, in which the Feller condition
fails. Every section is strictly decreasing, as \eqref{strictboundary}
asserts. For these parameters the boundary of set B lies above that of set A
at the same variance level, by at most about $0.04K$ on the plotted range,
and the two sets share every qualitative feature predicted by
Theorem~\ref{main}.}
\label{fig:slices}
\end{figure}

Figures~\ref{fig:surface} and \ref{fig:slices} show the boundary surface and
several of its sections. The extracted surface is nonincreasing in both
variables up to the resolution of the grid and shows no visible irregularity
for $\tau>0$ and $y>0$. Along the line $y=0$ the
computed boundary is finite and positive, which is consistent with the
threshold results of \cite{LTprop}, and Theorem~\ref{mainaxis}
establishes endpoint matching, bounded speed in maturity, and
$C^1$ regularity of every spatial slice at positive maturity. The ridge of
the surface at $\tau=0$, where every section starts at the strike,
illustrates Proposition~\ref{prop:expiry}. The displayed mesh is not
used to estimate an asymptotic gap exponent near the axis, and the
steepness of the sections at small variance in the right column of
Figure~\ref{fig:slices} is the qualitative counterpart of
$b_y\to-\infty$.

Formula \eqref{gamma} can be tested on the same output. Since $w$ is
smooth up to the boundary from the continuation side, the second difference
of $w$ in $x$ at the third to sixth continuation nodes was extrapolated
linearly back to $x=\beta$, which gives $w_{xx}^+$ and hence
$P_{ss}(\tau,b+,y)=w_{xx}^+/b^2$. The slope $b_y$ was obtained from a local
quadratic fit of the extracted boundary over a window of width $0.1$ in $y$,
because the boundary location carries an error below the grid size that a centered
difference over two neighboring nodes would amplify. Table~\ref{tab:gamma}
compares this estimate with the right-hand side of \eqref{gamma} evaluated
at the computed $b$ and $b_y$. For $y\ge0.09$ the two agree within $1.7\%$,
which is the accuracy one can expect from a first-order time discretization
and a boundary location read off the grid. At $y=0.04$ the boundary is steep in $y$
and the quadratic contact region is thin relative to the grid, and the
discrepancy grows to between $2.4\%$ and $3.2\%$. The table is an
illustration of the identity, not a convergence study.

\begin{table}[htbp]
\centering
\caption{Boundary gamma for parameter set A. The column $P_{ss}^\mathrm{fd}$ is
the finite difference estimate of the gamma from the continuation side at the
boundary, and the column $P_{ss}^\mathrm{formula}$ is \eqref{gamma} evaluated
with the computed $b$ and $b_y$.}
\label{tab:gamma}
\begin{tabular}{@{}ccccccc@{}}
\toprule
$\tau$ & $y$ & $b/K$ & $b_y/K$ & $P_{ss}^\mathrm{fd}$ & $P_{ss}^\mathrm{formula}$ & difference\\
\midrule
$0.25$ & $0.04$ & $0.8783$ & $-1.730$ & $4.125$ & $4.028$ & $+2.4\%$\\
$0.25$ & $0.09$ & $0.8098$ & $-1.179$ & $1.955$ & $1.977$ & $-1.2\%$\\
$0.25$ & $0.16$ & $0.7411$ & $-0.852$ & $1.238$ & $1.259$ & $-1.7\%$\\
$0.25$ & $0.36$ & $0.6135$ & $-0.496$ & $0.761$ & $0.774$ & $-1.7\%$\\
$0.25$ & $0.64$ & $0.5047$ & $-0.310$ & $0.627$ & $0.631$ & $-0.6\%$\\
$0.50$ & $0.04$ & $0.8520$ & $-1.785$ & $4.549$ & $4.407$ & $+3.2\%$\\
$0.50$ & $0.09$ & $0.7797$ & $-1.256$ & $2.264$ & $2.258$ & $+0.3\%$\\
$0.50$ & $0.16$ & $0.7057$ & $-0.917$ & $1.481$ & $1.485$ & $-0.3\%$\\
$0.50$ & $0.36$ & $0.5689$ & $-0.527$ & $0.958$ & $0.964$ & $-0.6\%$\\
$0.50$ & $0.64$ & $0.4539$ & $-0.323$ & $0.829$ & $0.831$ & $-0.3\%$\\
$1.00$ & $0.04$ & $0.8200$ & $-1.714$ & $4.977$ & $4.848$ & $+2.7\%$\\
$1.00$ & $0.09$ & $0.7496$ & $-1.233$ & $2.535$ & $2.505$ & $+1.2\%$\\
$1.00$ & $0.16$ & $0.6766$ & $-0.915$ & $1.676$ & $1.669$ & $+0.4\%$\\
$1.00$ & $0.36$ & $0.5390$ & $-0.533$ & $1.112$ & $1.115$ & $-0.3\%$\\
$1.00$ & $0.64$ & $0.4234$ & $-0.324$ & $0.992$ & $0.990$ & $+0.2\%$\\
\bottomrule
\end{tabular}
\end{table}

\noindent\textbf{Funding.} Junkee Jeon is supported by the National Research
Foundation of Korea (NRF) grant funded by the Korea government [Grant No.
RS-2026-25591924]. Takwon Kim is supported by the National Research
Foundation of Korea (NRF) grant funded by the Korea government [Grant No.
RS-2024-00351151]. Jehan Oh is supported by the National Research Foundation
of Korea (NRF) grant funded by the Korea government [Grant Nos.
RS-2025-00555316 and RS-2025-25415411].

\medskip\noindent\textbf{Data availability.} No empirical data were used in
this study, and all numerical parameters are illustrative. The code that
produced the numerical illustrations is available from the corresponding
author upon reasonable request. The boundary figures and the gamma table
illustrate Theorem~\ref{main} and are not numerical tests of the endpoint
theorems.

\providecommand{\bysame}{\leavevmode\hbox to3em{\hrulefill}\thinspace}
\providecommand{\MR}{\relax\ifhmode\unskip\space\fi MR }
\providecommand{\MRhref}[2]{%
  \href{http://www.ams.org/mathscinet-getitem?mr=#1}{#2}
}
\providecommand{\href}[2]{#2}


\begin{thebibliography}{10}

\bibitem{Alfonsi}
Aur\'elien Alfonsi, \emph{Affine diffusions and related processes: Simulation,
  theory and applications}, Bocconi \& Springer Series, vol.~6, Springer, Cham,
  2015.

\bibitem{AJO}
Sigurd Assing, Saul Jacka, and Adriana Ocejo, \emph{Monotonicity of the value
  function for a two-dimensional optimal stopping problem}, Ann. Appl. Probab.
  \textbf{24} (2014), 1554--1584.

\bibitem{AK}
Alessandro Audrito and Teo Kukuljan, \emph{Regularity theory for fully
  nonlinear parabolic obstacle problems}, J. Funct. Anal. \textbf{285} (2023),
  110116.

\bibitem{BDM}
Adrien Blanchet, Jean Dolbeault, and R\'egis Monneau, \emph{On the continuity
  of the time derivative of the solution to the parabolic obstacle problem with
  variable coefficients}, J. Math. Pures Appl. \textbf{85} (2006), 371--414,
  Erratum: J. Math. Pures Appl. 94 (2010), pp. 447--449.

\bibitem{Caf77}
Luis~A. Caffarelli, \emph{The regularity of free boundaries in higher
  dimensions}, Acta Math. \textbf{139} (1977), 155--184.

\bibitem{Caf98}
\bysame, \emph{The obstacle problem revisited}, J. Fourier Anal. Appl.
  \textbf{4} (1998), 383--402.

\bibitem{CPS}
Luis~A. Caffarelli, Arshak Petrosyan, and Henrik Shahgholian, \emph{Regularity
  of a free boundary in parabolic potential theory}, J. Amer. Math. Soc.
  \textbf{17} (2004), 827--869.

\bibitem{CDP}
Cheng Cai, Tiziano De~Angelis, and Jan Palczewski, \emph{The {A}merican put
  with finite-time maturity and stochastic interest rate}, Math. Finance
  \textbf{32} (2022), 1170--1213.

\bibitem{CC}
Xinfu Chen and John Chadam, \emph{A mathematical analysis of the optimal
  exercise boundary for {A}merican put options}, SIAM J. Math. Anal.
  \textbf{38} (2007), 1613--1641.

\bibitem{CCJZ}
Xinfu Chen, John Chadam, Lishang Jiang, and Weian Zheng, \emph{Convexity of the
  exercise boundary of the {A}merican put option on a zero dividend asset},
  Math. Finance \textbf{18} (2008), 185--197.

\bibitem{CJS}
Xinfu Chen, Zhou Jin, and Qingshuo Song, \emph{{A}merican option model and
  negative {F}ichera function on degenerate boundary}, Modeling, Stochastic
  Control, Optimization, and Applications, IMA Vol. Math. Appl., vol. 164,
  Springer, Cham, 2019, pp.~95--113.

\bibitem{CIR}
John~C. Cox, Jonathan~E. Ingersoll, and Stephen~A. Ross, \emph{A theory of the
  term structure of interest rates}, Econometrica \textbf{53} (1985), 385--407.

\bibitem{CIL}
Michael~G. Crandall, Hitoshi Ishii, and Pierre-Louis Lions, \emph{User's guide
  to viscosity solutions of second order partial differential equations}, Bull.
  Amer. Math. Soc. (N.S.) \textbf{27} (1992), 1--67.

\bibitem{DF}
Panagiota Daskalopoulos and Paul M.~N. Feehan, \emph{{$C^{1,1}$} regularity for
  degenerate elliptic obstacle problems}, J. Differential Equations
  \textbf{260} (2016), 5043--5074.

\bibitem{DP20}
Tiziano De~Angelis and Goran Peskir, \emph{Global {$C^1$} regularity of the
  value function in optimal stopping problems}, Ann. Appl. Probab. \textbf{30}
  (2020), 1007--1031.

\bibitem{Detemple}
J\'er\^ome Detemple, \emph{American-style derivatives: Valuation and
  computation}, Chapman \& Hall/CRC Financial Mathematics Series, Chapman \&
  Hall/CRC, Boca Raton, FL, 2006.

\bibitem{Ekstrom}
Erik Ekstr\"om, \emph{Properties of {A}merican option prices}, Stochastic
  Process. Appl. \textbf{114} (2004), 265--278.

\bibitem{FP}
Paul M.~N. Feehan and Camelia~A. Pop, \emph{A {S}chauder approach to
  degenerate-parabolic partial differential equations with unbounded
  coefficients}, J. Differential Equations \textbf{254} (2013), 4401--4445.

\bibitem{Feller}
William Feller, \emph{Two singular diffusion problems}, Ann. of Math. (2)
  \textbf{54} (1951), 173--182.

\bibitem{Friedman75}
Avner Friedman, \emph{Parabolic variational inequalities in one space dimension
  and smoothness of the free boundary}, J. Functional Analysis \textbf{18}
  (1975), 151--176.

\bibitem{Heston}
Steven~L. Heston, \emph{A closed-form solution for options with stochastic
  volatility with applications to bond and currency options}, Rev. Financ.
  Stud. \textbf{6} (1993), 327--343.

\bibitem{IW}
Nobuyuki Ikeda and Shinzo Watanabe, \emph{Stochastic differential equations and
  diffusion processes}, 2nd ed., North-Holland Mathematical Library, vol.~24,
  North-Holland, Amsterdam, 1989.

\bibitem{IT}
Samuli Ikonen and Jari Toivanen, \emph{Operator splitting methods for pricing
  {A}merican options under stochastic volatility}, Numer. Math. \textbf{113}
  (2009), 299--324.

\bibitem{Jacka}
Saul~D. Jacka, \emph{Optimal stopping and the {A}merican put}, Math. Finance
  \textbf{1} (1991), no.~2, 1--14.

\bibitem{JLL}
Patrick Jaillet, Damien Lamberton, and Bernard Lapeyre, \emph{Variational
  inequalities and the pricing of {A}merican options}, Acta Appl. Math.
  \textbf{21} (1990), 263--289.

\bibitem{KS}
Ioannis Karatzas and Steven~E. Shreve, \emph{Brownian motion and stochastic
  calculus}, 2nd ed., Graduate Texts in Mathematics, vol. 113, Springer, New
  York, 1991.

\bibitem{Kim}
In~Joon Kim, \emph{The analytic valuation of {A}merican options}, Rev. Financ.
  Stud. \textbf{3} (1990), 547--572.

\bibitem{Kuk}
Teo Kukuljan, \emph{Higher order parabolic boundary {H}arnack inequality in
  {$C^1$} and {$C^{k,\alpha}$} domains}, Discrete Contin. Dyn. Syst.
  \textbf{42} (2022), 2667--2698.

\bibitem{LSU}
Olga~A. Lady\v{z}enskaja, Vsevolod~A. Solonnikov, and Nina~N. Ural'ceva,
  \emph{Linear and quasilinear equations of parabolic type}, Translations of
  Mathematical Monographs, vol.~23, American Mathematical Society, Providence,
  RI, 1968.

\bibitem{LTprop}
Damien Lamberton and Giulia Terenzi, \emph{Properties of the {A}merican price
  function in the {H}eston-type models}, arXiv:1904.01653, 2019.

\bibitem{LTvi}
\bysame, \emph{Variational formulation of {A}merican option prices in the
  {H}eston model}, SIAM J. Financial Math. \textbf{10} (2019), 261--308.

\bibitem{LS}
Peter Laurence and Sandro Salsa, \emph{Regularity of the free boundary of an
  {A}merican option on several assets}, Comm. Pure Appl. Math. \textbf{62}
  (2009), 969--994.

\bibitem{Lieberman}
Gary~M. Lieberman, \emph{Second order parabolic differential equations}, World
  Scientific, River Edge, NJ, 1996.

\bibitem{Myneni}
Ravi Myneni, \emph{The pricing of the {A}merican option}, Ann. Appl. Probab.
  \textbf{2} (1992), 1--23.

\bibitem{NC}
Khai Nguyen and Huy Chau, \emph{On regularity of finite-maturity {A}merican put
  options in the {H}eston model}, arXiv:2606.10191, 2026.

\bibitem{Peskir05}
Goran Peskir, \emph{On the {A}merican option problem}, Math. Finance
  \textbf{15} (2005), 169--181.

\bibitem{Peskir19}
\bysame, \emph{Continuity of the optimal stopping boundary for two-dimensional
  diffusions}, Ann. Appl. Probab. \textbf{29} (2019), 505--530.

\bibitem{PS}
Goran Peskir and Albert Shiryaev, \emph{Optimal stopping and free-boundary
  problems}, Lectures in Mathematics ETH Z\"urich, Birkh\"auser, Basel, 2006.

\bibitem{PS07}
Arshak Petrosyan and Henrik Shahgholian, \emph{Parabolic obstacle problems
  applied to finance}, Recent Developments in Nonlinear Partial Differential
  Equations, Contemp. Math., vol. 439, American Mathematical Society,
  Providence, RI, 2007, pp.~117--133.

\bibitem{PSU}
Arshak Petrosyan, Henrik Shahgholian, and Nina Uraltseva, \emph{Regularity of
  free boundaries in obstacle-type problems}, Graduate Studies in Mathematics,
  vol. 136, American Mathematical Society, Providence, RI, 2012.

\bibitem{PitmanYor}
Jim Pitman and Marc Yor, \emph{A decomposition of {B}essel bridges}, Z.
  Wahrsch. Verw. Gebiete \textbf{59} (1982), 425--457.

\bibitem{RY}
Daniel Revuz and Marc Yor, \emph{Continuous martingales and brownian motion},
  3rd ed., Grundlehren der Mathematischen Wissenschaften, vol. 293, Springer,
  Berlin, 1999.

\bibitem{Touzi}
Nizar Touzi, \emph{{A}merican options exercise boundary when the volatility
  changes randomly}, Appl. Math. Optim. \textbf{39} (1999), 411--422.

\bibitem{WZ}
Lihe Wang and Yunhua Zhang, \emph{{$C^{1,\alpha}$} regularity for degenerate
  parabolic equations arising from the {H}eston model}, Commun. Pure Appl.
  Anal. \textbf{22} (2023), 3430--3460.

\bibitem{YW}
Toshio Yamada and Shinzo Watanabe, \emph{On the uniqueness of solutions of
  stochastic differential equations}, J. Math. Kyoto Univ. \textbf{11} (1971),
  155--167.

\end{thebibliography}
\end{document}